%% file: main.tex
\pdfoutput=1
\documentclass[sigconf,nonacm]{acmart}

\usepackage{multirow}
\usepackage{graphicx}
\usepackage{caption}
\usepackage{subcaption}

\usepackage{algorithm}
\usepackage[noend]{algpseudocode}

\begin{document}

\title[Enriching Simple Keyword Queries for Domain-Aware Narrative Retrieval]{Enriching Simple Keyword Queries for\\ Domain-Aware Narrative Retrieval}

\author{Hermann Kroll}
\email{kroll@ifis.cs.tu-bs.de}
\orcid{0000-0001-9887-9276}
\affiliation{%
  \institution{Institute for Information Systems, TU Braunschweig}
  \streetaddress{Mühlenpfordtstr. 23}
  \city{Braunschweig}
  \state{Lower Saxony}
  \country{Germany}
  \postcode{38106}
} 

\author{Christin Katharina Kreutz}
\email{christin.kreutz@th-koeln.de}
\orcid{0000-0002-5075-7699}
\affiliation{%
  \institution{
  TH Köln - University of Applied Sciences}
  \city{Cologne}
  \state{North Rhine-Westphalia}
  \country{Germany}
  \postcode{}
}  

\author{Pascal Sackhoff}
\email{p.sackhoff@tu-bs.de}
\affiliation{%
  \institution{Institute for Information Systems, TU Braunschweig}
  \streetaddress{Mühlenpfordtstr. 23}
  \city{Braunschweig}
  \state{Lower Saxony}
  \country{Germany}
  \postcode{38106}
}

\author{Wolf-Tilo Balke}
\email{balke@ifis.cs.tu-bs.de}
\orcid{0000-0002-5443-1215}
\affiliation{%
  \institution{Institute for Information Systems, TU Braunschweig}
  \streetaddress{Mühlenpfordtstr. 23}
  \city{Braunschweig}
  \state{Lower Saxony}
  \country{Germany}
  \postcode{38106}
}   

\renewcommand{\shortauthors}{Kroll et al.}

\begin{abstract}
Providing effective access paths to content is a key task in digital libraries. Oftentimes, such access paths are realized through advanced query languages, which, on the one hand, users may find challenging to learn or use, and on the other, requires libraries to convert their content into a high quality structured representation. As a remedy, narrative information access proposes to query library content through structured patterns directly, to ensure validity and coherence of information. However, users still find it challenging to express their information needs in such patterns. Therefore, this work bridges the gap by introducing a method that deduces patterns from keyword searches. Moreover, our user studies with participants from the biomedical domain show their acceptance of our prototypical system.

\end{abstract}



\maketitle

\section{Introduction}
\input{01-Introduction}

\section{Related Work}
\label{sec:relatedwork}
\input{02-RelatedWork}

\section{Query Model and Retrieval}
\label{sec:querymodel}
\input{03-QueryModel}

\section{Keywords to Narrative Queries}
\input{04-Algorithm.tex}

\section{System Implementation}

\input{05-Implementation.tex}

\section{User Studies}
\input{06-UserStudies}

\section{Evaluation of Effectiveness}

\input{07-AlgorithmEvaluation}

\section{Conclusions}
\input{08-Conclusion}

\section*{Acknowledgments}
Supported by the Deutsche Forschungsgemeinschaft (DFG, German Research Foundation): PubPharm – the Specialized Information Service for Pharmacy (Gepris 267140244).

\bibliographystyle{ACM-Reference-Format}
\bibliography{references}

\end{document}

%% file: 01-Introduction.tex
Digital libraries maintain extensive collections of scientific literature and make them accessible for a variety of uses. 
For search, generally simple yet intuitive keyword-based access paths are implemented, see, e.g., PubMed, Google Scholar, or dblp~\cite{DBLP:journals/pvldb/Ley09}. 
However, while such access paths are easy to use and relatively cheap to implement, users may benefit from more advanced access paths.
Here, using simple keyword-based search is oftentimes insufficient due to the limited expressiveness and the considerable amount of domain knowledge required to focus and/or refine searches~\cite{DBLP:journals/jodl/KreutzS22}.

As a remedy, user experience and retrieval quality can be severely boosted by advanced access paths paired with thorough semantic enrichment of digital library objects~\cite{10.1093/database/baq036}.
In particular, this means to annotate publications with domain-specific concepts and connecting relations. 
In this line, scientific knowledge bases (KBs)  provide innovative ways to access literature using advanced query types like navigational queries or graph patterns, e.g., KBs in \cite{2022openalex,faerber2019microsoftacademicknowledgegraph,DBLP:conf/acl/BettsPA19,jaradeh2019openknowledgeresearchgraph}. 
While navigational and exploratory queries are beneficial in practice, they require users to be proficient in complex query languages like SQL or SPARQL. 
As a remedy, keyword search on databases and knowledge bases has been proposed
~\cite{DBLP:conf/semweb/LiangPYZLM21,DBLP:conf/ercimdl/GkirtzouKVD15,10.1145/2063576.2063615,10.1016/j.websem.2009.07.005}. 
However, it is still challenging for digital libraries to convert their content into such structured representations with an acceptable quality, e.g., designing reliable extraction workflows. 

A different approach, the so-called narrative information access~\cite{kroll2022narrativeinformationaccess}, allows users to formulate their information needs as short narratives (stories) of interest -- involving relevant concepts and their interactions.
The main advantage is that narrative information access puts again textual publications of digital libraries in its focus. 
An example of such a querying mechanism are the so-called narrative query graphs, basically directed edge-labeled graph patterns~\cite{DBLP:conf/icadl/KrollPKKRB21}. 
Different from other knowledge base approaches, these patterns are matched against single publications instead of a single knowledge base to preserve validity and coherence of information~\cite{DBLP:conf/ercimdl/KrollKNMB20}. 
So, users can precisely retrieve suitable publications, and additionally, generate structured overviews of the literature. 
However, a query log analysis of their system revealed that formulating pattern-like queries is already challenging for users; See Sect.~\ref{sec:relatedwork}.
Our overall goal is to allow users to formulate their information needs as intuitively and easily as possible, i.e., as keyword queries, while providing a system capable of satisfying their possibly complex demands.  
Consequently, this work deals with the following research question \textbf{RQ}: \textit{Can a user's search intent be deduced from a keyword query?}

In this work, we strive to better connect users' actual information expression strategies with narrative information access, i.e., our proposed method translates  keyword to narrative queries. 
Unfortunately, deducing narrative queries from keywords might be ambiguous. 
We therefore propose a feedback loop: 
Users state keywords, we generate narrative queries, selection strategies select the best query options concerning different criteria, and the best queries are visualized for the users to choose from. 
However, integrating users into the process requires a suitable query representation, so that they can assess the generated patterns quickly and easily. 
We therefore perform user studies to answer the following questions: \textit{Q1. How should generated patterns be presented to the users, i.e., which query representation is suitable for our users? Q2. How useful is the end-to-end system?}

To answer the first question, we interviewed domain experts and asked them to complete a qualitative questionnaire. 
Still, one may ask whether our suggested workflow (keyword entering + graph generation + user selection) is suitable to support users. 
Thus we asked experts to utilize our prototypical system before we interviewed them to understand the usefulness and suitability of the end-to-end system.
But, \textit{how effectively does our method translate keyword-based queries to narrative queries for users (Q3)?}
So, we analyzed our method's quality on biomedical retrieval benchmarks to better generalize our findings, involving abstracts and full-texts, highly specific and more general queries, and natural language questions. 
Our contribution is thus an effective digital library system that is accepted by users in our domain.

%% file: 02-RelatedWork.tex
\textbf{PubPharm's Narrative Discovery System.}
In PubPharm, the German specialized information service for Pharmacy, we have implemented narrative information access since 2021\footnote{\url{http://narrative.pubpharm.de}}~\cite{kroll2022narrativeinformationaccess,DBLP:conf/ercimdl/KrollMB21}.
Our system allows users to formulate their information need as a graph query which is then matched against graph representations of biomedical articles~\cite{DBLP:conf/icadl/KrollPKKRB21}. 
In a pre-processing step, biomedical articles are converted into graph representations by identifying biomedical concepts and their relationships. 
Users can then search via a list of statements (basically concept-interactions) and the system replies with matching documents that contain the searched statements. 

However, a query log analysis from 2021 and 2022 revealed that only 440 of 7268 queries contained more than a single statement (concept interaction). 
This means that users refrain from formulating complex queries. 
Discussions with our users later revealed difficulties in formulating more complex query patterns. 
The triple-based query construction seemed not intuitive enough. 
In this work, we built upon our narrative retrieval system~\cite{Kroll2023IJDL} by simplifying users' interaction with the system while keeping its  expressiveness.

\textbf{Graph-based Retrieval.}
Dietz et al. proposed the usage of knowledge graphs for text-centric information retrieval~\cite{dietz2018kgfortextretrieval}. 
They discussed how entities, graph structures, and relations might be incorporated to boost retrieval quality.  
Another work discussed how open relation extraction might accelerate passage retrieval for given entities in queries~\cite{kadry2017openreforretrieval}. 
Although both works are related to ours, they rather suggest features and first evaluations instead of implementing a complete system. 
Krause~\cite{DBLP:phd/dnb/Krause19} developed a graph-based retrieval system making texts more accessible which differs from our user-focus.

\textbf{Pattern Mining.}
In general, pattern mining aims to find useful patterns in data, e.g., association rule mining discovers rules from existing data.
Mining rules in knowledge bases then allows to infer new facts (complete KBs), or finding errors~\cite{DBLP:conf/www/GalarragaTHS13}. 
Fang et al.~\cite{DBLP:journals/pvldb/FangSYB11} produce interesting explanations for connections between entity pairs in KBs by constrained graph patterns and path enumeration algorithms. 
Saleh and Pecina~\cite{DBLP:conf/ecir/SalehP19} propose query expansion for cross-lingual medical information retrieval while focusing on the vocabulary mismatch problem. 
In contrast, our work is focused on deducing narrative queries from keywords, visualizing them for users, and letting users search with them.

\textbf{Keyword Queries on Knowledge Bases.}
Searching KBs with keywords has already been explored~\cite{DBLP:conf/semweb/LiangPYZLM21,DBLP:conf/ercimdl/GkirtzouKVD15}.
Existing approaches~\cite{DBLP:conf/icde/BhalotiaHNCS02,DBLP:conf/ercimdl/BikakisGLSDS13,DBLP:conf/sigmod/HeWYY07} usually work as follows:
\textit{i)} map the keywords to structured data elements,
\textit{ii)} connect the keywords by searching for substructures
and \textit{iii)} rank the retrieved substructures via a scoring function. 
Gkirtzou et al.~\cite{DBLP:conf/ercimdl/GkirtzouKVD15} proposed keyword-based searches on RDF-type data sources. 
Elbassuoni and Blanco~\cite{10.1145/2063576.2063615} use a backtracking algorithm to construct RDF subgraphs from keyword queries to retrieve information from RDF graphs. 
All triples in a KB are treated as documents where the components (subject, predicate, object) represent its content. 
Maximum subgraphs are constructed by retrieving documents for each query keyword (producing \#keywords lists) and merging triples from different lists as subgraphs. 
They re-rank these subgraphs with statistical language models.
Zenz et al.~\cite{10.1016/j.websem.2009.07.005} propose QUICK, an RDF schema-based approach. 
While disregarding the actual keywords, in a first step they construct query templates from one-edged templates and recursively extend them by new edges. 
The second step associates the keywords from the query to the properties, classes and concepts from the templates.

While keyword search on knowledge bases is related to our work, the main difference is however, that the previous works in this domain assume a single KB with a known schema. 
In contrast, our data model includes millions of small documents graphs which allows a new definition of support, i.e., we can estimate how many graphs support a generated query.

%
\textbf{Natural Language Queries.}
Another area of research is based on directly stating queries in natural language and automatically translating them into query languages like SQL (known as NL2SQL).
Affolter et al.~\cite{DBLP:journals/vldb/AffolterSB19} present a survey comparing different textual query to database approaches and categorize them into keyword-, pattern-, parsing- and grammar-based, depending on their underlying methodology.  
Gkini et al.~\cite{10.1145/3448016.3452836} study text-to-SQL approaches from a performance point of view with their benchmarking system. 
Liang et al.~\cite{DBLP:conf/semweb/LiangPYZLM21} propose an end-to-end BERT-based model to transfer natural language queries into subject-relation-object triples. 
They also jointly learn the auxiliary tasks of output variable selection, query type classification and ordinal constraint detection.
Revanth et al.~\cite{10.1007/978-981-16-1342-5_21} transform natural language expressions in English to SQL queries by using NLP methods: lexical analysis, syntactic analysis, semantic analysis and transform the outputs. 
ChatGPT is one of the most recent methods.

The main limitation of these approaches is that they require training data to learn the actual translation. 
In practice, this can be an issue for digital libraries as typically not enough training data (natural language queries and their ideal translation) is available. 
That is why we decided to design an unsupervised translation method that does not require training data. 
To evaluate our algorithm on natural language questions, we selected a suitable biomedical benchmark.

\textbf{Query Visualization in Digital Libraries.}
Keyword-based retrieval systems have been well established for scientific information needs, e.g., PubMed, Google Scholar, Scopus, or dblp~\cite{DBLP:journals/pvldb/Ley09}. 
In brief, these systems typically capture the user's keywords in some text input field and display results as a list to the user.
When systems utilize semantic search techniques through machine learning, these systems typically boost the retrieval quality~\cite{10.1093/database/baq036,DBLP:journals/csur/TamineG22}.
However, they usually do not visualize what is happening with the query to the users, e.g., Semantic Scholar~\cite{ammar-etal-2018-construction}.
Knowledge bases like Wikidata~\cite{DBLP:conf/www/Vrandecic12} or the Open Research Knowledge Graph~\cite{jaradeh2019openknowledgeresearchgraph} provide the users either with entity-centric interfaces to click and navigate through the knowledge, or with SPARQL endpoints requiring users to learn SPARQL for posing queries.
In contrast, we enrich keyword queries and present derived queries in a feedback loop.

%% file: 03-QueryModel.tex
In our work~\cite{DBLP:conf/icadl/KrollPKKRB21}, we defined narrative query graphs as directed edge-labeled graphs with concepts as nodes and relationships as edges between them. 
However, we observed two major drawbacks of our system: 
First, users may not know how a certain concept should be connected to a query pattern. 
Suppose a user search for case-based studies in connection with Metformin diabetes treatments. 
In this case, a \texttt{treats} relationship can be placed between the concepts \texttt{Metformin} and \texttt{diabetes}. 
But to which concept should the concept \texttt{case-based studies} at best be connected? 
Second, some users faced the out-of-vocabulary problem, i.e., they wanted to search for concepts that were not known in the system. 
In this work we adjust our previous query model to tackle both drawbacks: In addition to graph patterns, we allow queries to also search for concepts that are not connected and for terms to tackle the out-of-vocabulary problem.

Formally, we denote $\mathcal{C}$ as the set of known concepts. 
Each \textbf{concept} $c$ is identified by an identifier, e.g., $c_{\textit{Metformin}}=(\textit{CHEMBL1431})$. 
Concepts are usually collected and arranged in domain-specific taxonomies or ontologies, e.g., the Medical Subject Headings\footnote{\url{http://meshb.nlm.nih.gov}}.
Some concepts may be arranged in a \textbf{subconcept} relation, e.g., \textit{Diabetes Mellitus Type 1} is sub concept of the super concept \textit{Diabetes Mellitus}. 
We denote relationships between two concepts by $\mathcal{R}$ -- the set of  predicates (also known as relationship labels), e.g., \texttt{associated} or \texttt{treats}.
Those predicates might be very general like \texttt{associated} or could be more specific like \texttt{treats}. 
In Wikidata for example, predicates are understood as resources/items that can be arranged in an hierarchy. 
Note that some domains might not arrange their predicates in this way. 
With that, we can define a \textbf{statement} as a triple, e.g.,  ($c_{\textit{Metformin}}$, \texttt{treats}, $c_{\textit{Diabetes Mellitus}}$).
Next, we define the set of statements as $\textit{Statements} \subseteq \mathcal{C} \times \mathcal{R} \times \mathcal{C}$.
Note that our definition of statements is similar to the representation of knowledge in the Resource Description Framework (RDF)~\cite{manola2004rdf}. 

The retrieval system returns documents as results.
Therefore, we denote $\mathcal{D}$ as the set of documents. 
Our documents $d \in \mathcal{D}$ consist of texts and each text consists of terms (single words/tokens). 
$\mathcal{T}$ is the set of all terms and $t \in \mathcal{T}$ is some term. 
For the actual retrieval, we need to know the terms of a document, which concepts have been detected in it, and which statements were extracted from it. 

\begin{enumerate}
    \item $\textit{doc\_terms}(d)  = \lbrace t \in \mathcal{T} \mid  t \textit{ is term in } d \rbrace$
    \item $\textit{doc\_concepts}(d) = \lbrace c \in \mathcal{C} \mid c \textit{ detected in } d\rbrace$
    \item $\textit{doc\_stmts}(d) = \lbrace s \in \textit{Statements} \mid s \textit{ extracted from } d\rbrace$
\end{enumerate}

Finally, we define a \textbf{narrative query} $nq = (Q_S, Q_C, Q_T)$ with $Q_S \subseteq \textit{Statements}$, $Q_C  \subseteq \mathcal{C}$ and $Q_T \subseteq \mathcal{T}$.
In other words, a query may ask for statements, concepts, and terms.
We call $d$ a \textbf{match} with regard to $nq = (Q_S, Q_C, Q_T)$, iff the following conditions hold:
1. $Q_S \subseteq \textit{doc\_stmts}(d)$,
2. $Q_C \subseteq \textit{doc\_concepts}(d)$, 
3. $Q_T \subseteq \textit{doc\_terms}(d)$. 
The set of all document matches regarding a query $nq$ can then be defined as 
$\textit{answers}(nq) = \lbrace d \in \mathcal{D} \mid d \textit{ is match to } nq\rbrace$.

Now we define the \textbf{query translation task} as:

\textit{Given a keyword query $q = (w_1, \ldots, w_n) $ with terms $w_i$, find all narrative queries such that each term of $q$ is mapped to some query term, query concept or query statement.
In other words, all words of keyword query $q$ must be reflected in some way.}

%% file: 04-Algorithm.tex
In this section, we present our algorithm to deduce narrative queries from keywords. 
Given a keyword query, our goal is therefore to first generate \textbf{all} possible narrative queries, i.e., all combinations that can be derived. 
In a second step, the \textbf{best} queries concerning different criteria are selected to be shown to the users. 
However, generating all possible narrative queries from keywords could yield a plethora of queries as keywords might refer to concepts, predicates, or, even worse, be synonymous with a set of concepts.
And then, we still would have to place predicates between those concepts to derive statements. 
We first introduce suitable lookup indexes to minimize the generation space. 
For simplicity, we call a query's terms, concepts, and statements \textbf{query components}.

We utilize a \textbf{document collection index} that retrieves \textbf{support} of query components, i.e., how many documents in our collection include the corresponding component. 
Possible query components with low support (e.g., no documents) could be disregarded because queries with those components will yield empty (or fewer) results in the end. 
We design this index as an inverted index. 
Given the functions $doc\_terms$, $doc\_concepts$, and $doc\_stmts$, we iterate over all documents of the collection, use the functions to retrieve the required entries and finally build this index. 

The set of predicates $\mathcal{R}$ is typically known, either it is known during the information extraction from texts, or it can be derived by iterating over all statements.
In brief, we design a \textbf{predicate index} that maps labels to predicates. 
Suppose a user may enter a keyword query such as: \texttt{Metformin therapy Diabetes Mellitus}.
In that case, it would be beneficial to detect that the keyword \texttt{therapy} refers to a \texttt{treats} predicate.
In practical knowledge bases, predicates are usually described by human-readable labels and sometimes by a list of synonyms, e.g., Wikidata includes a list of \textit{also known as} labels (Property P2175). 
If such information is available, we can additionally incorporate it in our predicate index.

Our primary goal is the generation of narrative queries with concepts and their statements, i.e., we need to map keywords to concepts.
To do so, each concept of $\mathcal{C}$ should, at best, be described by a human-readable label and a list of synonyms (e.g., Wikidata's \textit{also known as} labels). 
To align keywords with concepts, we utilize those labels (label + synonyms) to compute an \textbf{concept index} that maps labels to concepts.
In practice, homonyms might exist, i.e., a label could refer to a set of concepts and not only to a single one. 
In that case, the subsequent step generates different queries -- at least one for each homonymous concepts.

\subsection{Generating Narrative Queries}
In general, one could assume a certain order of keywords, i.e., a subsequent list of keywords is an information unit and corresponds to a concept.
However, users might extend or refine keyword queries which may break such an order.
Our algorithm provides the option to consider all keyword permutations when mapping. 
Another point to think about is that users may query with arbitrary, information-sparse keywords such as \textit{of} or \textit{in}.
So, we provide the option to ignore stopwords in users' queries. 
Our algorithm operates as follows:

\textit{1. Mapping Phase.}
The first step takes a list of keywords (the user's keyword query) as its input. 
We optionally remove stopwords and then tokenize those keywords into tokens.
This step yields all possible mappings from keywords to concepts, predicates, and terms.
We iterate over all tokens in the keyword query and check if a token maps to a concept.
Note that a concept/predicate label might also consist of multiple tokens. 
Thus we also need to check combinations of keywords in this mapping phase. 
By default we assume that the tokens in a query follow a certain order, and thus, only check combinations of subsequent tokens, e.g., in \textit{Metformin treats Diabetes Mellitus} we would check the following combinations \textit{Mellitus}, \textit{Diabetes Mellitus}, \textit{treats Diabetes Mellitus}, \textit{Metformin treats Diabetes Mellitus} but not check the combination \textit{Metformin Diabetes Mellitus}. 
The non-default option does consider permutations, i.e., also \textit{Metformin Diabetes Mellitus}.

Next our document collection index comes into play. 
For this check, we introduce $\tau$ as a threshold parameter (default 0). 
We remove all mappings to concepts having a support below $\tau$, i.e., each concept must at least be included in  more than $\tau$ documents. 
Having our concept mappings, we compute possible statements by iterating over all concept combinations $(c_i, c_j)$. 
An inner loop iterates over all predicates $\mathcal{R}$ with the variable $p$. 
We then test whether each statement $(c_i, p, c_j)$ has support $> \tau$ in our collection index.
If yes, we keep the possible statement, i.e., we store it in a list.
If not, we ignore the statement because it is not supported enough and will thus yield too few or no documents. 
The last step is to map tokens to possible predicates if the predicate label or one of its synonyms matches the token (or multiple tokens).
If so, we add the token to the predicate mapping. 
Similarly, predicate labels may consist of multiple terms, so we also check token combinations here as done for the concept mappings. 
Finally, this step yields 1) mappings from query tokens to concepts and to predicates, and 2) a list of possible statements.

\begin{figure*}
    \centering
    \includegraphics[width=0.7\textwidth]{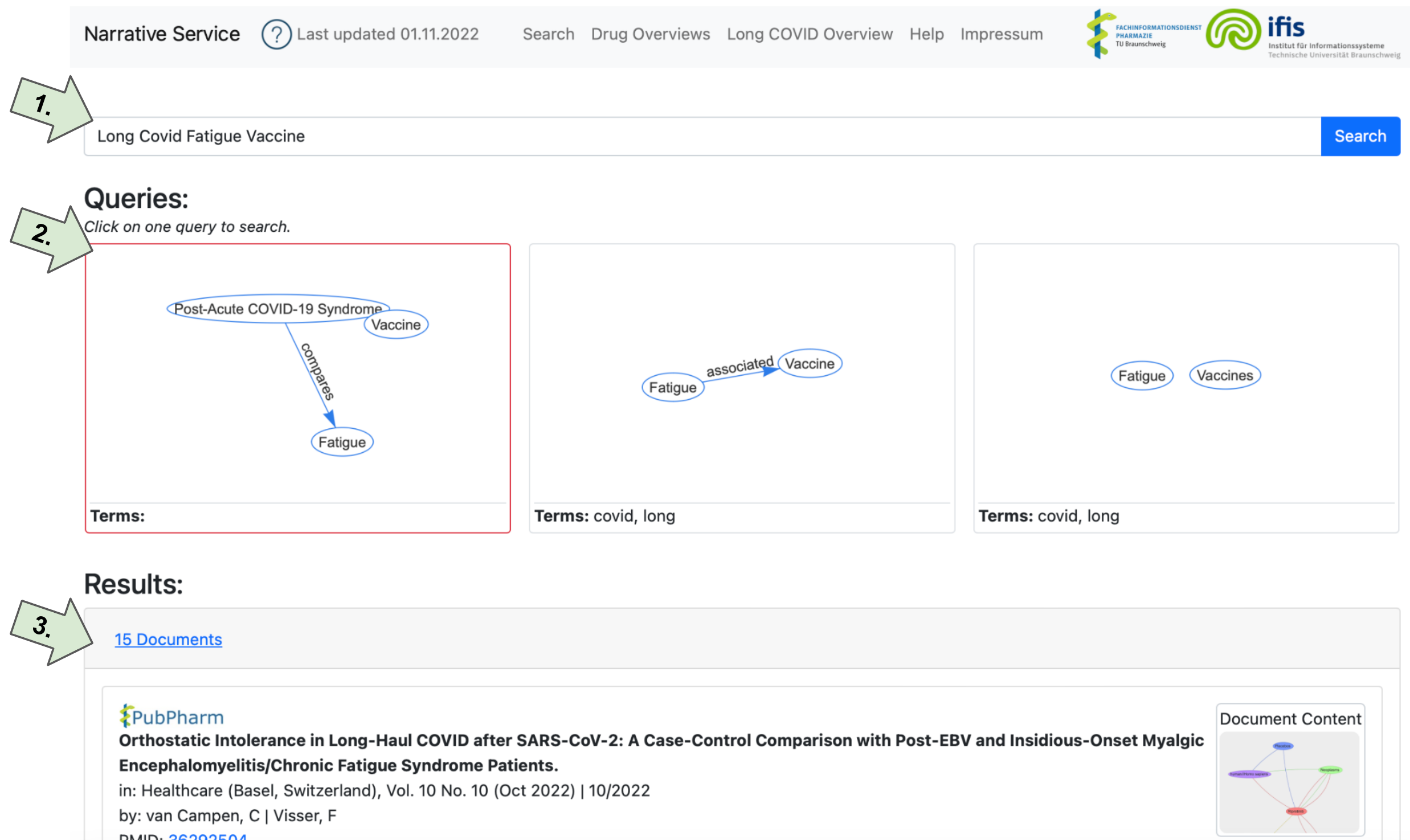}
    \caption{Prototypical user interface with markings. 1. highlights the search slit, 2. gives the query variants which users can choose from, 3. indicates the results of the query variant.}
    \label{fig:system_interface}
\end{figure*}

\textit{2. Generation Phase.}
The second step takes the query tokens, both mappings (concept and predicate), and a list of possible statements as input. 
The central strategy for this step is to map all tokens unambiguously to components of a narrative query. 
In other words, for each generated query, we must decide what we do with a token, i.e., whether we map it to a concept, a term, or a predicate, but not to multiple ones at the same time.
Furthermore, for each decision, we might have multiple options.
The algorithm works as follows:

\textbf{1. Map} tokens to concepts and predicates. If a token can be mapped to multiple concepts, generate a combination for each one. Also include the option not to map a token. This allows term-based only queries and every combination in between.

\textbf{2. Select Mappings.} For each combination from the previous step, generate a query. In this query, include the targets of the mappings (concepts and predicates). Also include those tokens that have not been mapped yet, as terms. Only keep those terms that have support $> \tau$. With this, we ensure that all tokens are mapped somehow, except token to term mappings that would yield too few document result.

\textbf{3. Integrate Statements.} For each query, we could decide which statements we include. Again, we have to compute all combinations here. That is why we compute a sub-list of statements from all possible statements (previous step) that applies to this query (the query must include the statement's concepts as concepts). Note, that we only allow putting a single predicate between two concepts. Then, compute all combinations (include a statement, do not include a statement). Again, generate queries for all different possibilities. 

\textbf{4. Filter} the generated queries with the following rule: If we map a keyword to a predicate, we must include a corresponding statement with that predicate in this query.  If no corresponding statement is included, the query is removed. 

Furthermore, we only allow putting a single predicate between two concepts because each query should represent a specific information need. 
If several predicates are possible, our algorithm  generates them as different queries. 
Note that checking all combinations generates a query where each keyword is mapped to a term.
We finally return a list of narrative queries. 
Note that the selection of $\tau$ will affect the overall exploration space since a low value forces our algorithm to generate more queries than a large value. 
However, for this paper, our goal was to generate all possible queries, i.e., we set $\tau = 0$.  
Including concepts, statements, or terms with a support $\leq \tau$ will not be helpful because all narrative queries asking for them will yield no or less results. 
Via our feedback loop in the user interface, we show which tokens have been excluded.
As an alternative, a system could include those tokens, return empty results, and force the users to refine their queries.

\subsection{Query Selection Strategies}
In the following, we introduce strategies to select the \textit{best} queries concerning different criteria. 
In brief, all strategies have to balance specificity and generality. 
We, therefore, design three strategies that we further analyze in this paper: a general one aiming for recall, a mixed one aiming for F1, and a specific one aiming for precision. 

The following two strategies should prefer queries with statements. 
That is why both strategies filter out all queries that do not contain a statement.  
Queries with statements can get very specific and may likely not yield any document results.
Due to our focus on users, each selected query should at least return some results. 
So, both strategies rank the queries with statements and only keep queries that yield at least a single result.
A predicate hierarchy may arrange predicates (see Sect.~\ref{sec:querymodel}), e.g.,  \texttt{treats} is more specific than \texttt{associated}, or \texttt{inhibits} is more specific than \texttt{induces}. 
In our biomedical use case, \textit{associated} is the most general predicate, and every other predicate is a specialization. 
In Wikidata, for instance, the \textit{Wikibase property} (Item Q29934218) could be seen as a very general predicate. 

That is why we designed the following two strategies:
The \textbf{mixed strategy} allows all predicates in queries.  
And the \textbf{specific strategy} forces queries to include specific predicates, i.e., prefers queries with more specialized predicates, e.g., prefers \texttt{treats} over \texttt{associated}. 
If predicates are not arranged in a hierarchy, the strategies select the same queries.  
As our motivation for these two strategies we assumed that selecting very specific predicates will likely boost the precision, but reduce the recall. 
A more general predicate might be a good mix between precision (because we force a statement) and recall (we do not force a to specific one). 
We still have to weigh the number of statements and the number of returned results. 
Due to our focus on users, we decided to rank the remaining queries by the number of results so that users can expect a fair number. 
Our last strategy focuses on recall. \textbf{The most-supported strategy} executes all generated queries and ranks the queries by the number of returned results in descending order, i.e., prefer queries that yield more documents than other queries.
Then the best ranked query is yielded. 
This strategy usually prefers term/concept-only queries because the number of hits is likely higher.

%% file: 05-Implementation.tex
We have already described implementation details for our narrative retrieval system in~\cite{kroll2022narrativeinformationaccess,DBLP:conf/icadl/KrollPKKRB21}. 
In the following, we implemented our algorithm and extended our previous retrieval system. 
The following numbers are based on a snapshot of the system from December 2022. 
The system worked on the whole biomedical National Library of Medicine (PubMed) Medline. 

We retrieved 35M documents (titles + abstracts), 711M concept annotations, and 842M extracted statements. 
The concepts stem from existing ontologies: the Medical Subject Headings, the ChEMBL database, and Wikidata~\cite{DBLP:conf/www/Vrandecic12}. 
The concept ontology had a root node (Thing) and then branched out into 13 basic concepts, e.g., drugs, diseases, targets, genes, species, etc. 
In sum, 635k concepts were known in the system.
The retrieval system organized ten different predicates into a hierarchy, e.g., \texttt{associated} is the most general predicate, and every predicate is a specialization of it. 
Information about the predicates (synonyms and hierarchy) can be found at\footnote{\url{http://www.narrative.pubpharm.de/help/}}.
Given the concepts and predicates plus synonyms, we derived our concept and predicate index. 
We then used the document data to compute our document collection index.

Consider some searches for diseases. 
In that case, all documents should support a \texttt{disease} concept if the disease concept or one of its subconcepts can be found.
We materialized, therefore, the concept ontology like suggested in~\cite{krotzsch2016your}, i.e., if a certain concept was found in a document, all super-concepts could also be found in that document. 
The same rule applies to statements:
Suppose the statement $(s, p, o)$ was extracted from some document.
In that case, we also store the same relation $p$ between all super-concept combinations of $s$ and $o$. 
In addition to that, we also store all statements with more general predicates, i.e., if a document contains a \texttt{treats} statement, it also implies a corresponding \texttt{associated} statement.

Next, we computed a case-insensitive inverted term index for those documents. 
To remove stopwords, we used the English NLTK stopword list~\cite{DBLP:conf/acl/Bird06}.
Therefore we iterated over the documents' contents, split the text by a space character, removed stopwords, and used the remaining tokens for indexing.
As an additional option we repeated that procedure but replaced all punctuation in texts with a space as biomedical concepts often contain special characters like \textit{-} and \textit{+}. 
We used the Python Punctuation set: !"\#\$\%\&'()*+,-./:;<=>?@[\textbackslash]\^\_`\{|\}~..
Finally, we computed the document collection index with 39M term and 635k concept, and 318M statement inverted index entries. 
For the query tokenizer, we removed brackets and split the keywords by a space.

Next, we implemented a prototypical user interface\footnote{\url{http://narrative.pubpharm.de/keyword_search/}} which is depicted in Figure~\ref{fig:system_interface}.
The interface works as follows:
1. Users enter a keyword query.
2. Three generated queries were visualized for the users.
3. A user's click on one of them started a search and returned matching documents.
We used our previously introduced strategies to select the three queries. 
If one strategy might not yield a query, e.g., we simply did not visualize it. 
Concerning the query visualization, we first conducted a user study which is described in the following section.
The study concluded that the graph representation was most suitable for our users.
We then implemented a graph representation for the second study, i.e., concepts were visualized as nodes, statements as edges between them and terms as a simple comma-separated list.

We used the same visualization for the document result lists as in the narrative retrieval service. 
Please note that one feature of this system is the support of variables in queries~\cite{DBLP:conf/icadl/KrollPKKRB21}.
A variable asks for any possible concept that fits into the query, e.g.,  for all \textit{diseases that can be treated by Metformin}. 
Internally, PubPharm rewrites queries that include some very general concepts like \texttt{disease}, \texttt{drug}, and \texttt{target} to aggregate the literature by suitable substations, i.e., showing one list of documents about the drug \texttt{Simvastatin} and one about \texttt{Metformin}.
For our first user study, we investigate whether we should include such information in our query representations, i.e., we included variables written as $?X$.

%% file: 06-UserStudies.tex
In the following, we describe our user studies and study results. 
For convenience, the user-centric evaluation was based on: \textit{Q1. How should generated patterns be presented to the users, i.e., which query representation is suitable for our users? Q2. How useful is the end-to-end system?}


\input{06.1_Study_I.tex}

\input{06.2_Study_II.tex}

\subsection{Discussion}

While participants of the first user study rated both the query and natural language representation as quite understandable at first glance (see Table~\ref{tab:res_understanable}), users' comments in the open questions and the group discussions showed their overall preference of the graph representation due to its clarity, ease of learning and capability of quickly conveying information. 
The second study supports this choice, participants mentioned the graph representations of queries being understood immediately.
Our users stated that pharmacists are usually experienced with graph representations, e.g., with visual chemical reactions or when drawing molecular structures. 
Our study revealed that the most suitable generated query patterns' representation for users  was the graph representation. 
We found all ten users of our second user study intuitively being able to use our prototypical end-to-end system.
Moreover, it was suitable for satisfying their individual information needs. 
Everyone confidently picked the graph pattern which best fit their query. 
However, we identified room for improvements regarding the different aspects of the system (see Section~\ref{sec:s2_results}) which should be rectified in a further iteration of the system.



%% file: 06.1_Study_I.tex
\subsection{Study I: Questionnaire and Discussion}
The overall goal of this study is to gain insights on query representations' suitability for (potential) users. 
This qualitative user study consists of two parts, a questionnaire and a following group discussion.
This study was conducted in context of an online workshop from PubPharm in which participants from the broader area of the pharmaceutical domain were introduced to PubPharm's narrative information access.
Participants took part in our study voluntarily.

\begin{figure}
    \centering
    \begin{subfigure}[b]{0.47\textwidth}
         \centering
         \includegraphics[width=0.9\textwidth]{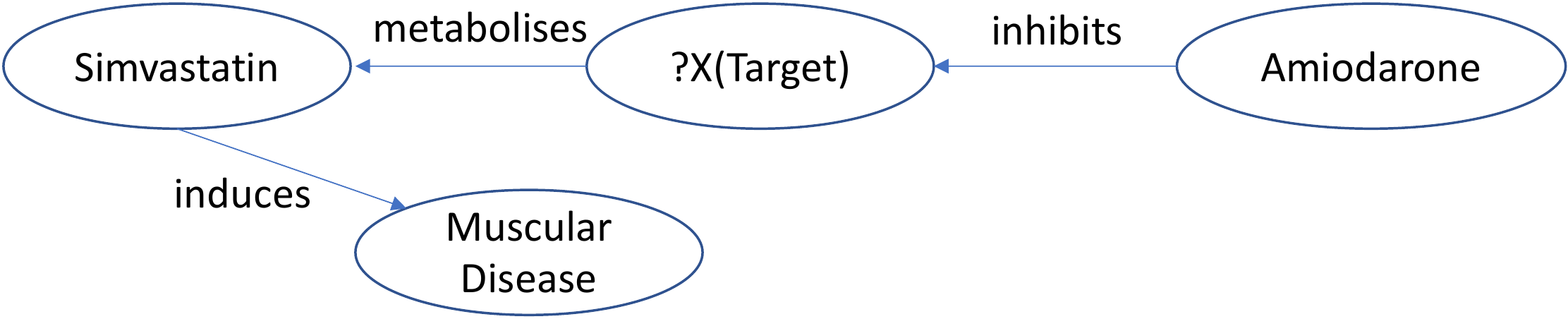}
         \caption{Graph query of \textit{'Through which target do Simvastatin and Amiodarone interact so that Simvastatin may induce a Muscular Disease?'}}
     \end{subfigure}
     
     \begin{subfigure}[b]{0.47\textwidth}
         \centering
         \includegraphics[width=0.9\textwidth]{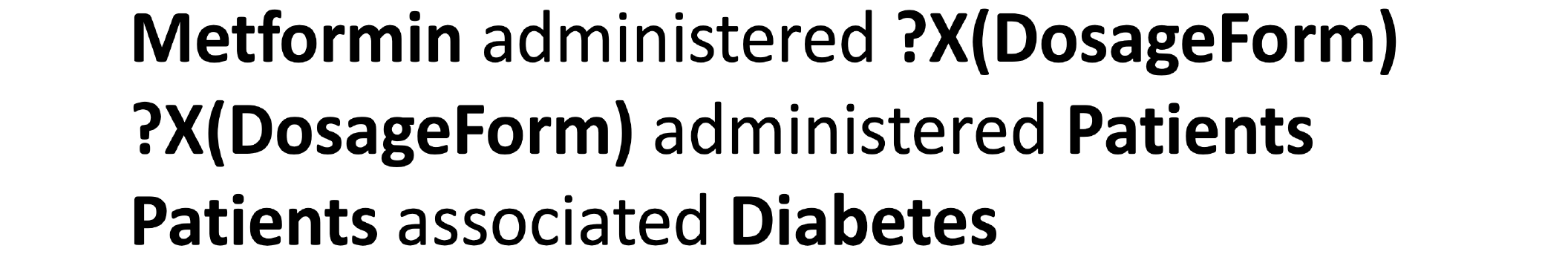}
         \caption{Structured query of \textit{'As which dosage forms can Metformin be administered to diabetic patients?'}}
     \end{subfigure}
     
     \begin{subfigure}[b]{0.47\textwidth}
         \centering
         \includegraphics[width=0.9\textwidth]{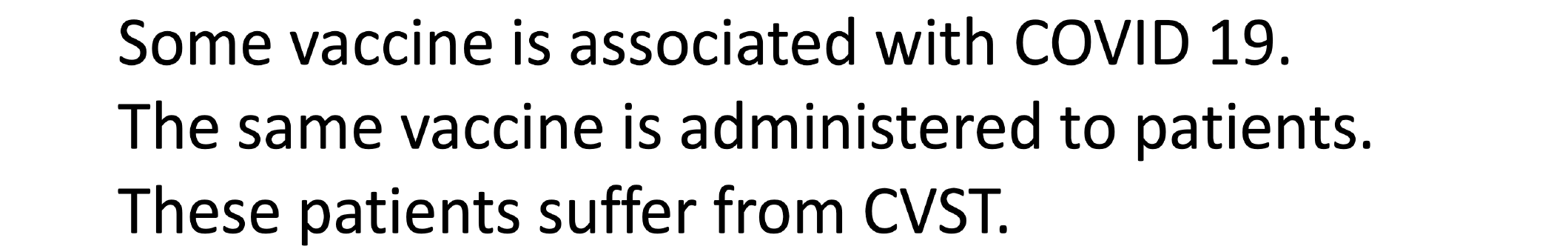}
        \caption{Natural language query of \textit{'Which COVID 19 vaccines may make patients suffer from cerebral venous sinus thrombosis (CVST)?'}}
     \end{subfigure}
    \caption{Query representations and information needs.}
    \label{fig:questionnaire_reps}
\end{figure}

\subsubsection{Setup - Questionnaire}
All study participants were presented an English online questionnaire. 
It first introduced three different query representations with different exemplary information needs.
We used these representations to have a mix of structured (graph), semi-structured (triple-like text statements) and natural language query representations (see Figure~\ref{fig:questionnaire_reps}).
We showed this exact order: graph, structured and natural language. 
The information needs were well-known examples from the biomedical domain and of the exact same structure. 
For each representation participants were asked if they immediately understood it on a 5-point Likert scale.

The next part of the questionnaire showed the three different information needs and possible representations (so the nine combinations) in one single figure. 
Study participants were then asked to answer (or skip) free text questions intending to capture their satisfaction with and the suitability of the representations. 
The \textit{questionnaire open questions} were derived from the main components of user satisfaction described in the user experience questionnaire\footnote{\url{https://www.ueq-online.org}}~\cite{ueq_paper}: 

\begin{itemize}
    \item[QQ$_1$] What did you like/dislike about the q. representations?
    \item[QQ$_2$] Which query representation would be/not be easy-to-learn for you and why?
    \item[QQ$_3$] Working with which query representation would/would not introduce unnecessary effort for you and why?
    \item[QQ$_4$] With which query representations would you be interested/ disinterested in working and why?
\end{itemize}

The first question QQ$_1$ strove to capture users' perception of \textit{attractiveness}, their overall impression and leaned onto the question from the UEQ, if users like or dislike a product (here the representation).
QQ$_2$ tackled \textit{perspicuity} and leaned on the UEQ's question if it is easy to get familiar with a product and to learn how to use it.
With QQ$_3$ we strove to observe \textit{efficiency}. 
This aspect of an UEQ usually assesses, if users can solve their tasks without unnecessary effort.
Lastly, QQ$_4$ aimed to look at \textit{stimulation}, so if users are excited and motivated to use a product.
We deliberately refrained from posing questions related to dependability and novelty of query representations in the questionnaire for time reasons. 
In our opinion dependability can only be assessed with actually using or constructing queries in the different query formulation.
Novelty is one of the less important factors in our case, as none of our query representations are truly novel. 
After these aspect-based open questions, participants were given the opportunity to answer what their favorite query representation was and to explain their choice.
Finally, the questionnaire asked them if they wanted to be contacted again for participating in another study on the same subject. 
For answering the whole questionnaire part, we gave our  participants 15 minutes.

\begin{table}[]
    \centering
    \caption{Study participants' ratings concerning the immediate understandability (IU) of the three compared query representations (++ strongly agree, -{}- strongly disagree).}
    \begin{tabular}{l|ccccc}
        \toprule
        IU & ++ & + & +/- & - & -{}-  \\ 
        \midrule
        graph & 3 & 3 & 2 & 1 & 0\\
        structured & 1 & 4 & 2 & 2 & 0\\
        natural language & 3 & 4 & 2 & 0 & 0\\
        \bottomrule
    \end{tabular}
    
    \label{tab:res_understanable}
\end{table}

\subsubsection{Results - Questionnaire: Likert Scales}
Nine participants answered the Likert Scale part of the evaluation by indicating their first impression on the understandability of the three query representations (see Table~\ref{tab:res_understanable}).
Study participants rated the graph representation and natural language representation similarly, the structured representation's immediate understandability was rated lower.

\subsubsection{Results - Questionnaire: Open Questions}
Usability aspects of the three query representation variants were evaluated through open questions by
seven of the nine initial participants:  

\textbf{QQ$_1$: Attractiveness.}
Participants indicated liked/disliked components regarding the representations:
%
They commented the graph representation was fast to understand in general.
It was taking time to be understood but was then stated as the best representation.
They stated that the graph was valuable because it visualizes relations, which play an important role in pharmacy. 
%
The structured representation was mentioned to take time to understand.
Two participants stated their unfamiliarity with its subject-predicate-object structure.
%
The natural language representation was mentioned as simple by three participants.
One mentioned it took time to understand while another one praised its quick understandability. 
One rated it as the best one. 
Another participant disliked its partially non-naturally sounding sentences. 
%
One participant disliked the absence of colorful highlighting in the graph representation.
Another one rated all representations as unintuitive.

\textbf{QQ$_2$: Perspicuity.}
Participants indicated the easiness of learning the representations:
%
Three participants rated the graph representation as the fastest to learn.
Two people mentioned it was easy to read and one mentioned it as the most complex representation.
%
The structured query representation was called confusing and unclear by one participant.
Another person commented it required some time to learn this representation.
%
One participant chose the natural language representation as the easiest to learn.
Another one labeled it as confusing and unclear.
%
In general one participant mentioned all representations being easy if one took the time to learn them while another participant rated them all as unintuitive.

\textbf{QQ$_3$: Efficiency.}
Participants indicated the level of unnecessary effort using the representations would introduce:
%
One person rated the graph representation as easy.
Another one mentioned it would be easy to learn.
%
The structured representation was seen as needing time to be learned. 
%
A participant deemed the natural language representation familiar. 
Another one found it imprecise and requiring more time to formulate.
%
Lastly, one person stated using any of the representation would not introduce unnecessary effort. 

\textbf{QQ$_4$: Stimulation.}
Participants indicated which representations they were interested/disinterested in using: 
%
Two mentioned the graph representation as positive while one participant rated the graph representation as the most complex one.
%
Another person disliked the structured representation as it would need to be learned first.
%
One participant commented that the natural language representation was the easiest to use.
%
Another person refrained from stating preferences in representations as these representations were merely a tool to answer interesting questions.

\textbf{Favorite.}
Five participants picked the graph representation as their favorite one and one picked the natural language representation. 
%
The graph representation's easiness and the possibility of visualizing complex interconnections were liked.

\subsubsection{Setup - Group Discussions}
The second part of this study were group discussions. 
Study participants were evenly divided into two groups (one in English, one in German) with an interviewer and a transcript writer each. 
The semi-structured group discussions took place directly after completing our questionnaire and took 10 minutes.
We asked them three \textit{guide questions} for the subsequent discussions:

\begin{itemize}
    \item[GQ$_1$] What representation could you imagine to use in practice?
    \item[GQ$_2$] What would you change? What should be different?
    \item[GQ$_3$] What was your favorite query representation and why?
\end{itemize}

\subsubsection{Results - Group Discussions}
Six study of the original nine participants took part in our group discussion. 
We evenly split them into two groups. 
In the following, we combine the opinions of both discussions.  
In each group, we started with three guiding questions. 

\textbf{GQ$_1$: Practical usage.}
One participant argued that if one is already using other information systems, they are used to a specific way of obtaining data. 
As all systems are different, simplifying usage and not introducing new query languages (so natural language representation) should be the focus. 
All other participants were more inclined towards the graphical representation. 
One participant stated that natural language seemed the easiest option in the beginning but was surpassed by the graph representation, as it clearly defines what is searched for. 
A further participant liked the graph representation, as it was easy to understand the query, but disliked the question marks in the representations (for the variables). 
They mentioned that the graph clearly shows the relations.
One interviewee stated the graphic representation would be the best one and all other representations would be very cumbersome. 
This view was shared by another study participant who additionally mentioned the graph representation would be easily practiced.
The natural language representation and structured representation with multiple rows was considered hard to read by multiple study participant.
However, one mentioned that the boldly marked concepts would be helpful to an extent.

\textbf{GQ$_2$: Desired changes.}
One participant considered the graph representation as not being self-explanatory and required the textual description of the information need to understand the representation. 
Another interviewee mentioned that users would need good examples to adapt them to their personal information needs. 
They further stated that with this help even complex graphs with more concepts could be constructed.
Someone suggested that arrows in the graph representation should be different from each other to convey information on the type of relation (e.g., distinguishing between a \textit{treats} and an \textit{inhibits} relation).
A study participant mentioned that (colorful) highlighting searched concepts could be helpful. 
Another one explicitly disliked having color in the graph representation as it would overload the query representation.

\textbf{GQ$_3$: Favorite.}
One of six participants preferred the natural language representation as it did not require a user to learn a new query language.
The remaining five participants preferred the graph representation because it would clearly highlight the connections between concepts and it would be easy to grasp.

%% file: 06.2_Study_II.tex
\subsection{Study II: Thinking-Aloud and Interview}
In the following, we analyze the usefulness of the end-to-end system for potential users. 
Our goal is to capture users' perspectives when deciding on a query pattern, their query formulation strategy, overall impression and problems with our prototypical system.
We therefore conducted a second user study fully online. 
It consisted of a thinking-aloud~\cite{thinking} exploration phase of our system (see Figure~\ref{fig:system_interface}) and a semi-structured interview. 
The sessions were conducted individually for each participant in the presence of two investigators.
In total the study took about 30 minutes per participant. 
Ten participants took part (nine in German, one in English).

We implemented the graph representation to visualize the queries as it was found to be the most suitable representation in our first user study.
As a small difference for better understandability, we adjusted the variable representation by removing the leading question mark and variable name, e.g., we replaced ?X(Drug) by Drug.

\subsubsection{Participants}
Our acquired participants were experts in the broader pharmaceutical domain and interested in the usage of domain-specific bibliographic information systems.
Those participants were researchers in the pharmaceutical domain in different positions (PhD students, postdocs, and professors). 
They took part in this study voluntarily and were explicitly made aware that they could refrain from taking part any time.

\subsubsection{Setup - Thinking-Aloud Exploration}
Before starting with the exploration, an investigator introduced the system (see Figure~\ref{fig:system_interface}) by showing a screenshot with instructions on how to use the system: 1. Enter a query, 2. Click on search, 3. Select a generated query, and 4. Explore the result list.

Participants fulfilled their own information needs from the domain with our prototypical system.
We gave them the guiding question: \textit{Think about the topic you are currently working on in the pharmaceutical domain. Which questions would you typically pose to PubMed or the keyword-based interface of PubPharm?} 
Participants were asked to describe their thoughts when interacting with the prototype system~\cite{thinking}. 
This step took 20 minutes at max.

\subsubsection{Setup - Semi-Structured Interview}
After the thinking-aloud exploration, semi-structured interviews with each participant tried to capture users' perspectives regarding the usefulness of the system. 
We used the following guide questions:

\begin{itemize}
    \item What are your general thoughts regarding the system?
    \item Where did you encounter problems? What was unclear?
    \item What did you like/immediately understand?
    \item Which changes would make you consider using t. system?
    \item Anything else you want to add or ask?
\end{itemize}

\subsubsection{Results}
\label{sec:s2_results}
This section discusses the observations, encountered problems and suggestions from the thinking-aloud exploration and the semi-structured interviews conjointly. 

In general, there were a lot of \textit{'This is what I meant'}-moments when graph patterns were generated for keyword queries. 
Displayed graph patterns were described to be immediately understood by participants.
The documents were found to be relevant when clicking on one of the graph patterns. 
Especially the combination of variables (e.g. diseases or targets) with concrete agents generated query patterns that were directly grasped by users. 

Users were very confident with choosing graph patterns fitting their query. 
Once they pick out a graph pattern, they do not select another one to compare the results. 
Out of all participants and queries only once a second graph pattern fitting the same keyword query was also chosen.
Beside the positive feedback, we found four core elements which lead to problems for most of the participants in varying degrees:

\textbf{Entering queries.}
For the query formulation, users suggested a prefix-based suggestion of concepts (autocompletion), support of multi part terms indicated by quotes and a spell correction of terms, e.g. entering \textit{pharmacokinetic} yielded no results but entering \textit{pharmacokinetics} yielded results. 

\textbf{Choosing a graph pattern.}
A graph pattern of a query should always be constructed. 
Study participants were confused by or displeased with query variants which only had terms or a combination of terms which were not visualized as graphs. 
These combinations seemed to not be intuitively understood by users. 
Additionally, more or better query variants should be displayed. 
Especially with a drug and a disease, only \textit{treats} and \textit{associated} were suggested as the connecting relations, while \textit{induces} would be another viable option to suggest.

\textbf{Filtering results.}
Filter options should be provided for users to navigate or restrict their results without having to change the query. 
Users seem to prefer restricting their current results opposed to writing narrower queries.
Participants of our study wanted to filter out documents by the year, the article type (e.g., surveys), and keywords contained in the title.

\textbf{Exploring results.}
The Provenance function (explains matches to users in the service) should be extended to include concepts and terms for users. 
Alternatively, the query part in the document content view needs to be better highlighted. 

\textbf{Further remarks.}
Multiple times study participants utilized the number of displayed results for a query to estimate if the query and the chosen graph pattern fit their information need. 
We therefore derive the approximate number of results being a valuable information which should be displayed in advance.

Other suggestions were: shorter loading times, graph patterns not overlapping, PubMed-like Boolean operator support in queries, visual structure search, a shopping-cart-style system to save interesting results and a direct integration of relevant results' citations/references which fit the query.

%% file: 07-AlgorithmEvaluation.tex
Our user studies have demonstrated that the graph representation was suitable and the overall system was accepted.
However, \textit{how effectively does our method translate keyword-based queries to narrative queries for users (Q3)?}

Due to our restriction to the biomedical domain, we had to restrict the evaluation to biomedical information retrieval benchmarks.
We picked the following ones:

\begin{enumerate}
\item \textbf{TREC Precision Medicine 2020}~\cite{TREC_PM20} (PM2020, 31 topics) covers precision-focused retrieval of biomedical PubMed abstracts. Each query asks for three concepts: a treatment (drug), a disease, and a variant (gene/target).
\item \textbf{TREC COVID 2020}~\cite{Voorhees2020TrecCovid} (COVID, 50 topics) bases on retrieval of COVID-relevant literature, the COVID Open Research Challenge~\cite{Wang2020Cord19}. We use the data's 5th (most recent, from 16th July '20) release. Queries ask for different topics on COVID-19, e.g., treatments, outbreaks.
\item \textbf{TREC Genomics 2007}~\cite{Hersh2007TrecGenomics} (Genom., 36 topics) includes natural language questions around biomedical target interactions for passage retrieval in full-texts.
\item \textbf{TripJudge}~\cite{althammer2022tripjudge} (TripJ., 1136 topics) holds queries and interaction data from the Trip Database for abstract retrieval. TripJudge is an extension of the TripClick~\cite{rekabsaz2021fairnessir} by improving the quality through human annotations.
\end{enumerate}

We choose these benchmarks to cover a wide range of biomedical queries, from very specific ones (PM2020) to general queries in TripJudge, up to natural language questions in Genomics. 
Titles and abstracts of the relevant documents were available for all benchmarks. 
Topics were single input queries.
PM2020 and TripJudge were based on abstract retrieval.
COVID could be evaluated in two ways: only abstracts and abstracts + full-texts.
Genomics, in contrast, was a full-text passage retrieval benchmark, and thus, we only evaluated the full-text setting. 
PubMed Medline documents required for PM2020 were already included in PubPharm's narrative system.
For the missing TripJudge, COVID (pre-prints), and Genomics documents (especially for the full-texts), we applied the same concept linking and information extraction, which has already been conducted for the PubMed collection.

\textbf{Setup.} Retrieval benchmarks typically provide judged documents, queries, and a ranking of which documents are relevant for a specific query. 
Usually, retrieval is evaluated by ranking documents and computing scores for different rank values $k$, like precision@$k$ and recall@$k$. 
However, our query model does Boolean retrieval, i.e., a document can be relevant for a query or not, there is no ranking among relevant documents.
Therefore, we had to compute the number of retrieved relevant documents and how many relevant documents were missing, i.e., we report precision, recall, and F1. 
For our subsequent evaluation, we follow the definition of bpref~\cite{10.1145/1277741.1277756} and only considered documents that have been judged in those benchmarks to determine whether a hit was relevant.

We designed our evaluation as follows:
First, we translated the queries of each benchmark into all possible narrative queries with our algorithm. 
We then executed those queries and took the ones that achieved the highest precision, recall, and F1 score for each query in every benchmark. 
This determined an upper bound on achievable precision, recall, and F1 with our retrieval model.
As a baseline, we used a Boolean term-based retrieval model, i.e., we used the terms of the queries directly for searching documents. 
Our second step then analyzed our selection strategies, i.e., we counted how many cases we selected one of the best precision queries, best-recall queries, and best-F1 queries. 
Note that multiple narrative queries may return the same score for some keyword queries. 
We can quantify how effectively our selection strategies pick the best queries concerning our evaluation metrics.

\begin{table}[t]
\centering
\caption{Highest-achievable retrieval quality with our query model compared to term-based retrieval.}

\begin{tabular}{l|l|c|ccc}
\toprule
& \textbf{Metric} & \textbf{TermB} & \textbf{BestP.} & \textbf{BestR.} & \textbf{BestF1}\\ \toprule

\multicolumn{6}{c}{\textbf{Abstract-Only Retrieval}} \\
\midrule
\multirow{3}{*}{\rotatebox[origin=c]{90}{\textbf{PM2020}}}& Prec. & 0.48 & \textbf{0.84} & 0.51 & 0.54 \\
& Rec. & 0.24 & 0.06 & \textbf{0.41} & 0.40 \\
& F1 & 0.27 & 0.10 & 0.40 & \textbf{0.41} \\

\midrule
\multirow{3}{*}{\rotatebox[origin=c]{90}{\textbf{COVID}}}& Prec. & 0.33 & \textbf{0.40} & 0.33 & 0.34 \\
& Rec. & 0.26 & 0.20 & \textbf{0.31} & 0.29 \\
& F1 & 0.22 & 0.17 & \textbf{0.24} & \textbf{0.24} \\

\midrule
\multirow{3}{*}{\rotatebox[origin=c]{90}{\textbf{TripJ.}}}& Prec. & 0.44 & \textbf{0.51} & 0.44 & 0.47 \\
& Rec. & 0.85 & 0.75 & \textbf{0.87} & 0.85 \\
& F1 & 0.53 & 0.50 & 0.53 & \textbf{0.55} \\

\midrule
\multicolumn{6}{c}{\textbf{Full-text Retrieval}}\\

\midrule
\multirow{3}{*}{\rotatebox[origin=c]{90}{\textbf{COVID}}}& Prec. & 0.16 & \textbf{0.26} & 0.16 & 0.18 \\
& Rec. & 0.45 & 0.32 & \textbf{0.49} & 0.44 \\
& F1 & 0.18 & 0.14 & 0.19 & \textbf{0.21} \\

\midrule
\multirow{3}{*}{\rotatebox[origin=c]{90}{\textbf{Genom.}}}& Prec. & 0.23 & \textbf{0.42} & 0.23 & 0.30 \\
& Rec. & 0.23 & 0.11 & \textbf{0.26} & 0.20 \\
& F1 & 0.14 & 0.12 & 0.14 & \textbf{0.18} \\

\bottomrule               
\end{tabular}

\label{tab:modelquality}
\end{table}

\begin{table}[t]
\centering
\caption{Number of topics where a query producing the highest metric was selected by one of our strategies. \textbf{'Any'} denotes where any of the best metrics queries was selected.}

\begin{tabular}{lr|rrr|r}
\toprule
\textbf{Benchmark} & \textbf{|Q|} & \textbf{BestP.} & \textbf{BestR.} & \textbf{BestF1} & \textbf{Any}\\ \toprule

\multicolumn{5}{c}{\textbf{Exact Query Found}} \\
\midrule
\textbf{PM2020} & 31 & 4 & 21 & 10 & 22 \\
\textbf{COVID} & 50 & 20 & 40 & 34 & 40 \\
\textbf{TripJ.} & 1136 & 548  & 806 & 579 & 849 \\
\midrule
\textbf{COVID+F} & 50 & 22 & 45 & 29 & 45 \\
\textbf{Genom.} & 36 & 12 & 23 & 16 & 25 \\
\midrule
\multicolumn{5}{c}{\textbf{One Allowed Edit in Terms/Concepts}} \\
\midrule
\textbf{PM2020} & 31 & 12 & 24 & 20 & 25 \\
\textbf{COVID} & 50 & 36 & 46 & 43 & 46 \\
\textbf{TripJ.} & 1136 & 804 & 926 & 839 & 947 \\
\midrule
\textbf{COVID+F} & 50 & 38 & 50 & 45 & 50 \\
\textbf{Genom.} & 36 & 16 & 30 & 23 & 30 \\

\midrule
\multicolumn{5}{c}{\textbf{One Allowed Edit in Predicates}} \\
\midrule
\textbf{PM2020} & 31 & 16 & 21 & 10 & 25 \\
\textbf{COVID} & 50 & 21 & 40 & 34 & 41 \\
\textbf{TripJ.} & 1136 & 569 & 807 & 594 & 854 \\
\midrule
\textbf{COVID+F} & 50 & 25 & 46 & 32 & 46 \\
\textbf{Genom.} & 36 & 15 & 24 & 18 & 27 \\

\bottomrule               
\end{tabular}

\label{tab:effectiveness}
\end{table}

\textbf{Effectiveness of Strategies.} 
Table~\ref{tab:modelquality} lists the best results when generating all narrative queries and using our query model for retrieval.
The evaluation showed that our query model improved the term-based search for every benchmark and metric.
For example, on TripJudge, the term-based retrieval achieved a precision of 0.44 and recall of 0.85, whereas our model boosted the precision to 0.51 with a recall of 0.75, or the recall to 0.87 by retaining the precision of 0.44.
The difference was even larger for PM2020, where the precision was boosted from 0.48 to 0.84 (by decreasing the recall from 0.24 to 0.06). 
Here, focusing on F1 boosted it from 0.27 to 0.41. 
Moreover, natural language questions of Genomics were translated into narrative queries that outperformed the baseline.

Further, we analyzed how many of those best queries contained statements and, thus, fully utilized our query model. 
We counted the number of topics for which the best query contained at least a single statement: 
Concerning precision, 26 out of 31 (PM2020), 12 out of 50 (COVID on full-texts), 195 out of 1136 (TripJudge), and 15 out of 36 topics (Genomics) did.
Concerning recall, 2 out of 31 (PM2020), 0 out of 50 (COVID on full-texts), 54 out of 1136 (TripJudge), and 6 out of 36 topics (Genomics) did.
Concerning F1, 3 out of 31 (PM2020), 5 out of 50 (COVID on full-texts), 85 out of 1136 (TripJudge), and 12 out of 36 (Genomics) did.
We expected these numbers because precision-oriented queries may rather contain statements than recall-oriented ones.
In addition, many benchmark queries, especially in TripJudge or COVID, were relatively short, e.g., \textit{coronavirus origin}, \textit{coronavirus quarantine}, or \textit{green tea}. 
Such keywords were not converted into statement-based queries because we only placed statements if we found at least two concepts. 
Here, our concept vocabulary did not include concepts for \textit{origin} and \textit{quarantine}. 
We counted how many of the best queries included at least two different concepts.
Concerning the queries with the best precision, 27 out of 31 (PM2020), 15 out of 50 (COVID on full-texts), 205 out of 1136 (TripJudge), and 17 out of 36 topics (Genomics) did. 
To verify that this was not just based on an out-of-concept-vocabulary problem, we counted how many queries contained three or more keywords: 31 out of 31 (PM2020), 36 out of 50 (COVID), 559 out of 1136 (TripJudge), and 36 out of 36 topics (Genomics) did.  
This justified our assumption that many benchmark topics were relatively short, i.e., we did not find concepts and, thus, did not ask for highly complex interactions. 
If queries tended to get longer and more precise, like in PM2020 or Genomics, statements in queries became more relevant. 

\textbf{Highest Metric Queries.} 
The evaluation demonstrated that our query model is indeed beneficial for information retrieval.
However, how many of those best queries do our selection strategies find in practice? 
In other words, what can users expect when entering different keyword queries? 
To answer this, we counted how often our three strategies yield one of the best possible queries concerning an evaluation metric (best precision, etc.). 
Table~\ref{tab:effectiveness} lists the results. 
For the 31 topics of PM2020, our strategies found the best precision queries for four topics, the best recall queries in 21 topics, the best F1 queries in ten topics, or at least one of the three best metric queries in 22 topics. 
For the 1136 topics of TripJudge, we found a best precision query for 548 topics, a best recall query for 806, a best F1 query for 579 topics, or at least one of the three best metric queries for 849 topics.
For the 36 Genomics natural language topics, for twelve topics our strategies found of the best precision, for 23 topics the the best recall, in 16 topics the best F1 one, or at least one of the best metric queries in 25 topics. 
In summary, users can expect to get the best query concerning precision in 13\% (PM2020) to 48\% (TripJudge) of cases. 
For best recall, they can expect queries in 67.7\% (PM2020) and 90\% (COVID on full-texts) of the cases.

\textbf{Allowed Edits.} 
However, how different are our selected queries if we do not find the best possible one?
We counted two cases: 
1. Queries that differ just in one term and concept, i.e., one query contains a keyword as a term, whereas the other query had it as a concept (one allowed edit in terms/concepts). 
2. Queries with the same statements except different predicates (one allowed edit in predicates). 
Of course, in both cases, those queries may differ concerning our metrics. 
However, it helped us to estimate how close our selected queries were compared to the best ones.   
The counts are reported in Table~\ref{tab:effectiveness}. 
Especially for PM2020, which had the lowest number of correctly selected queries concerning precision, we found twelve queries that just differed by one term/concept and 16 queries that had a different predicate. 
This finding also applied to the other benchmarks: Allowing a small edit in terms/concepts or a different predicate led to considerably better results.

\textbf{Discussion.}
First, our query model boosted the search for complex information needs, like stated in Genomics or PM2020. 
Next, our selection strategies did produce a high number of best queries concerning different evaluation metrics. 
And moreover, our methods were not adjusted for different benchmarks and were, thus, generalizable to a broad range of biomedical information needs.

%% file: 08-Conclusion.tex
In this work, we bridged the gap between the ease of keyword search and sophisticated narrative retrieval. 
Our evaluation has demonstrated that our proposed solutions were effective and generalized to a broad range of biomedical information needs.
Moreover, user studies with domain experts verified the usefulness of our prototypical system.
Especially from a digital library perspective, this work can be seen as a deep dive into how keyword-based search combined with sophisticated retrieval can be implemented, and, which possible challenges have to be faced on this way. 
Future work could design more advanced translation and selection strategies, improve the user interface based on our users' suggestions, and finally, investigate users' exploration strategies in a broader study to identify more requirements for the system.

%% file: main.bbl

\begin{thebibliography}{43}


\ifx \showCODEN    \undefined \def \showCODEN     #1{\unskip}     \fi
\ifx \showDOI      \undefined \def \showDOI       #1{#1}\fi
\ifx \showISBNx    \undefined \def \showISBNx     #1{\unskip}     \fi
\ifx \showISBNxiii \undefined \def \showISBNxiii  #1{\unskip}     \fi
\ifx \showISSN     \undefined \def \showISSN      #1{\unskip}     \fi
\ifx \showLCCN     \undefined \def \showLCCN      #1{\unskip}     \fi
\ifx \shownote     \undefined \def \shownote      #1{#1}          \fi
\ifx \showarticletitle \undefined \def \showarticletitle #1{#1}   \fi
\ifx \showURL      \undefined \def \showURL       {\relax}        \fi
\providecommand\bibfield[2]{#2}
\providecommand\bibinfo[2]{#2}
\providecommand\natexlab[1]{#1}
\providecommand\showeprint[2][]{arXiv:#2}

\bibitem[Affolter et~al\mbox{.}(2019)]%
        {DBLP:journals/vldb/AffolterSB19}
\bibfield{author}{\bibinfo{person}{Katrin Affolter}, \bibinfo{person}{Kurt
  Stockinger}, {and} \bibinfo{person}{Abraham Bernstein}.}
  \bibinfo{year}{2019}\natexlab{}.
\newblock \showarticletitle{{A comparative survey of recent natural language
  interfaces for databases}}.
\newblock \bibinfo{journal}{\emph{{VLDB} J.}} \bibinfo{volume}{28},
  \bibinfo{number}{5} (\bibinfo{year}{2019}), \bibinfo{pages}{793--819}.
\newblock
\urldef\tempurl%
\url{https://doi.org/10.1007/s00778-019-00567-8}
\showDOI{\tempurl}


\bibitem[Althammer et~al\mbox{.}(2022)]%
        {althammer2022tripjudge}
\bibfield{author}{\bibinfo{person}{Sophia Althammer},
  \bibinfo{person}{Sebastian Hofst\"{a}tter}, \bibinfo{person}{Suzan Verberne},
  {and} \bibinfo{person}{Allan Hanbury}.} \bibinfo{year}{2022}\natexlab{}.
\newblock \showarticletitle{{TripJudge: A Relevance Judgement Test Collection
  for TripClick Health Retrieval}}. In \bibinfo{booktitle}{\emph{Proceedings of
  the 31st ACM International Conference on Information and Knowledge
  Management}} (Atlanta, GA, USA) \emph{(\bibinfo{series}{CIKM '22})}.
  \bibinfo{publisher}{Association for Computing Machinery},
  \bibinfo{pages}{3801–3805}.
\newblock
\showISBNx{9781450392365}
\urldef\tempurl%
\url{https://doi.org/10.1145/3511808.3557714}
\showDOI{\tempurl}


\bibitem[Ammar et~al\mbox{.}(2018)]%
        {ammar-etal-2018-construction}
\bibfield{author}{\bibinfo{person}{Waleed Ammar}, \bibinfo{person}{Dirk
  Groeneveld}, \bibinfo{person}{Chandra Bhagavatula}, \bibinfo{person}{Iz
  Beltagy}, \bibinfo{person}{Miles Crawford}, \bibinfo{person}{Doug Downey},
  {and} \bibinfo{person}{al.}} \bibinfo{year}{2018}\natexlab{}.
\newblock \showarticletitle{{Construction of the Literature Graph in Semantic
  Scholar}}. In \bibinfo{booktitle}{\emph{Proceedings of the 2018 Conference of
  the North {A}merican Chapter of the Association for Computational
  Linguistics: Human Language Technologies, Volume 3 (Industry Papers)}}.
  \bibinfo{publisher}{Association for Computational Linguistics},
  \bibinfo{pages}{84--91}.
\newblock
\urldef\tempurl%
\url{https://doi.org/10.18653/v1/N18-3011}
\showDOI{\tempurl}


\bibitem[Betts et~al\mbox{.}(2019)]%
        {DBLP:conf/acl/BettsPA19}
\bibfield{author}{\bibinfo{person}{Christine Betts}, \bibinfo{person}{Joanna
  Power}, {and} \bibinfo{person}{Waleed Ammar}.}
  \bibinfo{year}{2019}\natexlab{}.
\newblock \showarticletitle{{GrapAL: Connecting the Dots in Scientific
  Literature}}. In \bibinfo{booktitle}{\emph{Proceedings of the 57th Conference
  of the Association for Computational Linguistics, {ACL} 2019, Florence,
  Italy, July 28 - August 2, 2019, Volume 3: System Demonstrations}}.
  \bibinfo{publisher}{Association for Computational Linguistics},
  \bibinfo{pages}{147--152}.
\newblock
\urldef\tempurl%
\url{https://doi.org/10.18653/v1/p19-3025}
\showDOI{\tempurl}


\bibitem[Bhalotia et~al\mbox{.}(2002)]%
        {DBLP:conf/icde/BhalotiaHNCS02}
\bibfield{author}{\bibinfo{person}{Gaurav Bhalotia}, \bibinfo{person}{Arvind
  Hulgeri}, \bibinfo{person}{Charuta Nakhe}, \bibinfo{person}{Soumen
  Chakrabarti}, {and} \bibinfo{person}{S. Sudarshan}.}
  \bibinfo{year}{2002}\natexlab{}.
\newblock \showarticletitle{{Keyword Searching and Browsing in Databases using
  {BANKS}}}. In \bibinfo{booktitle}{\emph{Proceedings of the 18th International
  Conference on Data Engineering, San Jose, CA, USA, February 26 - March 1,
  2002}}. \bibinfo{publisher}{{IEEE} Computer Society},
  \bibinfo{pages}{431--440}.
\newblock
\urldef\tempurl%
\url{https://doi.org/10.1109/ICDE.2002.994756}
\showDOI{\tempurl}


\bibitem[Bikakis et~al\mbox{.}(2013)]%
        {DBLP:conf/ercimdl/BikakisGLSDS13}
\bibfield{author}{\bibinfo{person}{Nikos Bikakis}, \bibinfo{person}{Giorgos
  Giannopoulos}, \bibinfo{person}{John Liagouris}, \bibinfo{person}{Dimitrios
  Skoutas}, \bibinfo{person}{Theodore Dalamagas}, {and}
  \bibinfo{person}{Timos~K. Sellis}.} \bibinfo{year}{2013}\natexlab{}.
\newblock \showarticletitle{{RDivF: Diversifying Keyword Search on {RDF}
  Graphs}}. In \bibinfo{booktitle}{\emph{Research and Advanced Technology for
  Digital Libraries - International Conference on Theory and Practice of
  Digital Libraries, {TPDL} 2013, Valletta, Malta, September 22-26, 2013.
  Proceedings}} \emph{(\bibinfo{series}{Lecture Notes in Computer Science},
  Vol.~\bibinfo{volume}{8092})}. \bibinfo{publisher}{Springer},
  \bibinfo{pages}{413--416}.
\newblock
\urldef\tempurl%
\url{https://doi.org/10.1007/978-3-642-40501-3\_49}
\showDOI{\tempurl}


\bibitem[Bird(2006)]%
        {DBLP:conf/acl/Bird06}
\bibfield{author}{\bibinfo{person}{Steven Bird}.}
  \bibinfo{year}{2006}\natexlab{}.
\newblock \showarticletitle{{{NLTK:} The Natural Language Toolkit}}. In
  \bibinfo{booktitle}{\emph{{ACL} 2006, 21st International Conference on
  Computational Linguistics and 44th Annual Meeting of the Association for
  Computational Linguistics, Proceedings of the Conference, Sydney, Australia,
  17-21 July 2006}}. \bibinfo{publisher}{The Association for Computer
  Linguistics}.
\newblock
\urldef\tempurl%
\url{https://doi.org/10.3115/1225403.1225421}
\showDOI{\tempurl}


\bibitem[Dietz et~al\mbox{.}(2018)]%
        {dietz2018kgfortextretrieval}
\bibfield{author}{\bibinfo{person}{Laura Dietz}, \bibinfo{person}{Alexander
  Kotov}, {and} \bibinfo{person}{Edgar Meij}.} \bibinfo{year}{2018}\natexlab{}.
\newblock \showarticletitle{{Utilizing Knowledge Graphs for Text-Centric
  Information Retrieval}}. In \bibinfo{booktitle}{\emph{The 41st International
  {ACM} {SIGIR} Conference on Research {\&} Development in Information
  Retrieval, {SIGIR} 2018, Ann Arbor, MI, USA, July 08-12, 2018}}.
  \bibinfo{publisher}{{ACM}}, \bibinfo{pages}{1387--1390}.
\newblock
\urldef\tempurl%
\url{https://doi.org/10.1145/3209978.3210187}
\showDOI{\tempurl}


\bibitem[Elbassuoni and Blanco(2011)]%
        {10.1145/2063576.2063615}
\bibfield{author}{\bibinfo{person}{Shady Elbassuoni} {and} \bibinfo{person}{Roi
  Blanco}.} \bibinfo{year}{2011}\natexlab{}.
\newblock \showarticletitle{{Keyword Search over RDF Graphs}}. In
  \bibinfo{booktitle}{\emph{Proceedings of the 20th ACM International
  Conference on Information and Knowledge Management}} (Glasgow, Scotland, UK)
  \emph{(\bibinfo{series}{CIKM '11})}. \bibinfo{publisher}{Association for
  Computing Machinery}, \bibinfo{pages}{237–242}.
\newblock
\showISBNx{9781450307178}
\urldef\tempurl%
\url{https://doi.org/10.1145/2063576.2063615}
\showDOI{\tempurl}


\bibitem[Fang et~al\mbox{.}(2011)]%
        {DBLP:journals/pvldb/FangSYB11}
\bibfield{author}{\bibinfo{person}{Lujun Fang}, \bibinfo{person}{Anish~Das
  Sarma}, \bibinfo{person}{Cong Yu}, {and} \bibinfo{person}{Philip Bohannon}.}
  \bibinfo{year}{2011}\natexlab{}.
\newblock \showarticletitle{{REX:} Explaining Relationships between Entity
  Pairs}.
\newblock \bibinfo{journal}{\emph{Proc. {VLDB} Endow.}} \bibinfo{volume}{5},
  \bibinfo{number}{3} (\bibinfo{year}{2011}), \bibinfo{pages}{241--252}.
\newblock
\urldef\tempurl%
\url{https://doi.org/10.14778/2078331.2078339}
\showDOI{\tempurl}


\bibitem[F{\"{a}}rber(2019)]%
        {faerber2019microsoftacademicknowledgegraph}
\bibfield{author}{\bibinfo{person}{Michael F{\"{a}}rber}.}
  \bibinfo{year}{2019}\natexlab{}.
\newblock \showarticletitle{{The Microsoft Academic Knowledge Graph: {A} Linked
  Data Source with 8 Billion Triples of Scholarly Data}}. In
  \bibinfo{booktitle}{\emph{The Semantic Web - {ISWC} 2019 - 18th International
  Semantic Web Conference, Auckland, New Zealand, October 26-30, 2019,
  Proceedings, Part {II}}} \emph{(\bibinfo{series}{Lecture Notes in Computer
  Science}, Vol.~\bibinfo{volume}{11779})}. \bibinfo{publisher}{Springer},
  \bibinfo{pages}{113--129}.
\newblock
\urldef\tempurl%
\url{https://doi.org/10.1007/978-3-030-30796-7\_8}
\showDOI{\tempurl}


\bibitem[Gal{\'{a}}rraga et~al\mbox{.}(2013)]%
        {DBLP:conf/www/GalarragaTHS13}
\bibfield{author}{\bibinfo{person}{Luis~Antonio Gal{\'{a}}rraga},
  \bibinfo{person}{Christina Teflioudi}, \bibinfo{person}{Katja Hose}, {and}
  \bibinfo{person}{Fabian~M. Suchanek}.} \bibinfo{year}{2013}\natexlab{}.
\newblock \showarticletitle{{AMIE:} association rule mining under incomplete
  evidence in ontological knowledge bases}. In \bibinfo{booktitle}{\emph{22nd
  International World Wide Web Conference, {WWW} '13, Rio de Janeiro, Brazil,
  May 13-17, 2013}}. \bibinfo{publisher}{International World Wide Web
  Conferences Steering Committee / {ACM}}, \bibinfo{pages}{413--422}.
\newblock
\urldef\tempurl%
\url{https://doi.org/10.1145/2488388.2488425}
\showDOI{\tempurl}


\bibitem[Gkini et~al\mbox{.}(2021)]%
        {10.1145/3448016.3452836}
\bibfield{author}{\bibinfo{person}{Orest Gkini}, \bibinfo{person}{Theofilos
  Belmpas}, \bibinfo{person}{Georgia Koutrika}, {and} \bibinfo{person}{Yannis
  Ioannidis}.} \bibinfo{year}{2021}\natexlab{}.
\newblock \showarticletitle{{An In-Depth Benchmarking of Text-to-SQL Systems}}.
  In \bibinfo{booktitle}{\emph{Proceedings of the 2021 International Conference
  on Management of Data}} (Virtual Event, China) \emph{(\bibinfo{series}{SIGMOD
  '21})}. \bibinfo{publisher}{Association for Computing Machinery},
  \bibinfo{pages}{632–644}.
\newblock
\showISBNx{9781450383431}
\urldef\tempurl%
\url{https://doi.org/10.1145/3448016.3452836}
\showDOI{\tempurl}


\bibitem[Gkirtzou et~al\mbox{.}(2015)]%
        {DBLP:conf/ercimdl/GkirtzouKVD15}
\bibfield{author}{\bibinfo{person}{Katerina Gkirtzou}, \bibinfo{person}{Kostis
  Karozos}, \bibinfo{person}{Vasilis Vassalos}, {and} \bibinfo{person}{Theodore
  Dalamagas}.} \bibinfo{year}{2015}\natexlab{}.
\newblock \showarticletitle{{Keywords-To-SPARQL Translation for {RDF} Data
  Search and Exploration}}. In \bibinfo{booktitle}{\emph{Research and Advanced
  Technology for Digital Libraries - 19th International Conference on Theory
  and Practice of Digital Libraries, {TPDL} 2015, Pozna{\'{n}}, Poland,
  September 14-18, 2015. Proceedings}} \emph{(\bibinfo{series}{Lecture Notes in
  Computer Science}, Vol.~\bibinfo{volume}{9316})}.
  \bibinfo{publisher}{Springer}, \bibinfo{pages}{111--123}.
\newblock
\urldef\tempurl%
\url{https://doi.org/10.1007/978-3-319-24592-8\_9}
\showDOI{\tempurl}


\bibitem[He et~al\mbox{.}(2007)]%
        {DBLP:conf/sigmod/HeWYY07}
\bibfield{author}{\bibinfo{person}{Hao He}, \bibinfo{person}{Haixun Wang},
  \bibinfo{person}{Jun Yang}, {and} \bibinfo{person}{Philip~S. Yu}.}
  \bibinfo{year}{2007}\natexlab{}.
\newblock \showarticletitle{{{BLINKS:} ranked keyword searches on graphs}}. In
  \bibinfo{booktitle}{\emph{Proceedings of the {ACM} {SIGMOD} International
  Conference on Management of Data, Beijing, China, June 12-14, 2007}}.
  \bibinfo{publisher}{{ACM}}, \bibinfo{pages}{305--316}.
\newblock
\urldef\tempurl%
\url{https://doi.org/10.1145/1247480.1247516}
\showDOI{\tempurl}


\bibitem[Hersh et~al\mbox{.}(2007)]%
        {Hersh2007TrecGenomics}
\bibfield{author}{\bibinfo{person}{William~R. Hersh}, \bibinfo{person}{Aaron~M.
  Cohen}, \bibinfo{person}{Lynn Ruslen}, {and} \bibinfo{person}{Phoebe~M.
  Roberts}.} \bibinfo{year}{2007}\natexlab{}.
\newblock \showarticletitle{{{TREC} 2007 Genomics Track Overview}}. In
  \bibinfo{booktitle}{\emph{Proceedings of The Sixteenth Text REtrieval
  Conference, {TREC} 2007, Gaithersburg, Maryland, USA, November 5-9, 2007}}
  \emph{(\bibinfo{series}{{NIST} Special Publication},
  Vol.~\bibinfo{volume}{500-274})}. \bibinfo{publisher}{National Institute of
  Standards and Technology {(NIST)}}.
\newblock
\urldef\tempurl%
\url{http://trec.nist.gov/pubs/trec16/papers/GEO.OVERVIEW16.pdf}
\showURL{%
\tempurl}


\bibitem[Jaradeh et~al\mbox{.}(2019)]%
        {jaradeh2019openknowledgeresearchgraph}
\bibfield{author}{\bibinfo{person}{Mohamad~Yaser Jaradeh},
  \bibinfo{person}{Allard Oelen}, \bibinfo{person}{Kheir~Eddine Farfar},
  \bibinfo{person}{Manuel Prinz}, \bibinfo{person}{Jennifer D'Souza},
  \bibinfo{person}{G{\'{a}}bor Kismih{\'{o}}k}, \bibinfo{person}{Markus
  Stocker}, {and} \bibinfo{person}{S{\"{o}}ren Auer}.}
  \bibinfo{year}{2019}\natexlab{}.
\newblock \showarticletitle{{Open Research Knowledge Graph: Next Generation
  Infrastructure for Semantic Scholarly Knowledge}}. In
  \bibinfo{booktitle}{\emph{Proceedings of the 10th International Conference on
  Knowledge Capture, {K-CAP} 2019, Marina Del Rey, CA, USA, November 19-21,
  2019}}. \bibinfo{publisher}{{ACM}}, \bibinfo{pages}{243--246}.
\newblock
\urldef\tempurl%
\url{https://doi.org/10.1145/3360901.3364435}
\showDOI{\tempurl}


\bibitem[Kadry and Dietz(2017)]%
        {kadry2017openreforretrieval}
\bibfield{author}{\bibinfo{person}{Amina Kadry} {and} \bibinfo{person}{Laura
  Dietz}.} \bibinfo{year}{2017}\natexlab{}.
\newblock \showarticletitle{{Open Relation Extraction for Support Passage
  Retrieval: Merit and Open Issues}}. In \bibinfo{booktitle}{\emph{Proceedings
  of the 40th International {ACM} {SIGIR} Conference on Research and
  Development in Information Retrieval, Shinjuku, Tokyo, Japan, August 7-11,
  2017}}. \bibinfo{publisher}{{ACM}}, \bibinfo{pages}{1149--1152}.
\newblock
\urldef\tempurl%
\url{https://doi.org/10.1145/3077136.3080744}
\showDOI{\tempurl}


\bibitem[Krause(2019)]%
        {DBLP:phd/dnb/Krause19}
\bibfield{author}{\bibinfo{person}{Thomas Krause}.}
  \bibinfo{year}{2019}\natexlab{}.
\newblock \emph{\bibinfo{title}{{{ANNIS:} {A} graph-based query system for
  deeply annotated text corpora}}}.
\newblock \bibinfo{thesistype}{Ph.\,D. Dissertation}. \bibinfo{school}{Humboldt
  University of Berlin, Germany}.
\newblock
\urldef\tempurl%
\url{http://edoc.hu-berlin.de/18452/20436}
\showURL{%
\tempurl}


\bibitem[Kreutz and Schenkel(2022)]%
        {DBLP:journals/jodl/KreutzS22}
\bibfield{author}{\bibinfo{person}{Christin~Katharina Kreutz} {and}
  \bibinfo{person}{Ralf Schenkel}.} \bibinfo{year}{2022}\natexlab{}.
\newblock \showarticletitle{{Scientific paper recommendation systems: a
  literature review of recent publications}}.
\newblock \bibinfo{journal}{\emph{Int. J. Digit. Libr.}} \bibinfo{volume}{23},
  \bibinfo{number}{4} (\bibinfo{year}{2022}), \bibinfo{pages}{335--369}.
\newblock
\urldef\tempurl%
\url{https://doi.org/10.1007/s00799-022-00339-w}
\showDOI{\tempurl}


\bibitem[Kroll et~al\mbox{.}(2020)]%
        {DBLP:conf/ercimdl/KrollKNMB20}
\bibfield{author}{\bibinfo{person}{Hermann Kroll},
  \bibinfo{person}{Jan{-}Christoph Kalo}, \bibinfo{person}{Denis Nagel},
  \bibinfo{person}{Stephan Mennicke}, {and} \bibinfo{person}{Wolf{-}Tilo
  Balke}.} \bibinfo{year}{2020}\natexlab{}.
\newblock \showarticletitle{{Context-Compatible Information Fusion for
  Scientific Knowledge Graphs}}. In \bibinfo{booktitle}{\emph{Digital Libraries
  for Open Knowledge - 24th International Conference on Theory and Practice of
  Digital Libraries, {TPDL} 2020, Lyon, France, August 25-27, 2020,
  Proceedings}} \emph{(\bibinfo{series}{Lecture Notes in Computer Science},
  Vol.~\bibinfo{volume}{12246})}. \bibinfo{publisher}{Springer},
  \bibinfo{pages}{33--47}.
\newblock
\urldef\tempurl%
\url{https://doi.org/10.1007/978-3-030-54956-5\_3}
\showDOI{\tempurl}


\bibitem[Kroll et~al\mbox{.}(2022a)]%
        {DBLP:conf/ercimdl/KrollMB21}
\bibfield{author}{\bibinfo{person}{Hermann Kroll}, \bibinfo{person}{Niklas
  Mainzer}, {and} \bibinfo{person}{Wolf{-}Tilo Balke}.}
  \bibinfo{year}{2022}\natexlab{a}.
\newblock \showarticletitle{{On Dimensions of Plausibility for Narrative
  Information Access to Digital Libraries}}. In
  \bibinfo{booktitle}{\emph{Linking Theory and Practice of Digital Libraries -
  26th International Conference on Theory and Practice of Digital Libraries,
  {TPDL} 2022, Padua, Italy, September 20-23, 2022, Proceedings}}
  \emph{(\bibinfo{series}{Lecture Notes in Computer Science},
  Vol.~\bibinfo{volume}{13541})}. \bibinfo{publisher}{Springer},
  \bibinfo{pages}{433--441}.
\newblock
\urldef\tempurl%
\url{https://doi.org/10.1007/978-3-031-16802-4\_43}
\showDOI{\tempurl}


\bibitem[Kroll et~al\mbox{.}(2021)]%
        {DBLP:conf/icadl/KrollPKKRB21}
\bibfield{author}{\bibinfo{person}{Hermann Kroll}, \bibinfo{person}{Jan
  Pirklbauer}, \bibinfo{person}{Jan{-}Christoph Kalo}, \bibinfo{person}{Morris
  Kunz}, \bibinfo{person}{Johannes Ruthmann}, {and}
  \bibinfo{person}{Wolf{-}Tilo Balke}.} \bibinfo{year}{2021}\natexlab{}.
\newblock \showarticletitle{{Narrative Query Graphs for
  Entity-Interaction-Aware Document Retrieval}}. In
  \bibinfo{booktitle}{\emph{Towards Open and Trustworthy Digital Societies -
  23rd International Conference on Asia-Pacific Digital Libraries, {ICADL}
  2021, Virtual Event, December 1-3, 2021, Proceedings}}
  \emph{(\bibinfo{series}{Lecture Notes in Computer Science},
  Vol.~\bibinfo{volume}{13133})}. \bibinfo{publisher}{Springer},
  \bibinfo{pages}{80--95}.
\newblock
\urldef\tempurl%
\url{https://doi.org/10.1007/978-3-030-91669-5\_7}
\showDOI{\tempurl}


\bibitem[Kroll et~al\mbox{.}(2023)]%
        {Kroll2023IJDL}
\bibfield{author}{\bibinfo{person}{Hermann Kroll}, \bibinfo{person}{Jan
  Pirklbauer}, \bibinfo{person}{Jan-Christoph Kalo}, \bibinfo{person}{Morris
  Kunz}, \bibinfo{person}{Johannes Ruthmann}, {and} \bibinfo{person}{Wolf-Tilo
  Balke}.} \bibinfo{year}{2023}\natexlab{}.
\newblock \showarticletitle{A discovery system for narrative query graphs:
  entity-interaction-aware document retrieval}.
\newblock \bibinfo{journal}{\emph{International Journal on Digital Libraries}}
  (\bibinfo{year}{2023}).
\newblock
\showISSN{1432-1300}
\urldef\tempurl%
\url{https://doi.org/10.1007/s00799-023-00356-3}
\showDOI{\tempurl}


\bibitem[Kroll et~al\mbox{.}(2022b)]%
        {kroll2022narrativeinformationaccess}
\bibfield{author}{\bibinfo{person}{Hermann Kroll}, \bibinfo{person}{Florian
  Pl\"{o}tzky}, \bibinfo{person}{Jan Pirklbauer}, {and}
  \bibinfo{person}{Wolf-Tilo Balke}.} \bibinfo{year}{2022}\natexlab{b}.
\newblock \showarticletitle{{What a Publication Tells You—Benefits of
  Narrative Information Access in Digital Libraries}}. In
  \bibinfo{booktitle}{\emph{Proceedings of the 22nd ACM/IEEE Joint Conference
  on Digital Libraries}} (Cologne, Germany) \emph{(\bibinfo{series}{JCDL
  '22})}. \bibinfo{publisher}{Association for Computing Machinery}, Article
  \bibinfo{articleno}{9}, \bibinfo{numpages}{8}~pages.
\newblock
\showISBNx{9781450393454}
\urldef\tempurl%
\url{https://doi.org/10.1145/3529372.3530928}
\showDOI{\tempurl}


\bibitem[Kr{\"o}tzsch and Rudolph(2016)]%
        {krotzsch2016your}
\bibfield{author}{\bibinfo{person}{Markus Kr{\"o}tzsch} {and}
  \bibinfo{person}{Sebastian Rudolph}.} \bibinfo{year}{2016}\natexlab{}.
\newblock \showarticletitle{{Is your database system a semantic web reasoner?}}
\newblock \bibinfo{journal}{\emph{KI-K{\"u}nstliche Intelligenz}}
  \bibinfo{volume}{30}, \bibinfo{number}{2} (\bibinfo{year}{2016}),
  \bibinfo{pages}{169--176}.
\newblock
\urldef\tempurl%
\url{https://doi.org/10.1007/s13218-015-0412-x}
\showDOI{\tempurl}


\bibitem[Laugwitz et~al\mbox{.}(2008)]%
        {ueq_paper}
\bibfield{author}{\bibinfo{person}{Bettina Laugwitz}, \bibinfo{person}{Theo
  Held}, {and} \bibinfo{person}{Martin Schrepp}.}
  \bibinfo{year}{2008}\natexlab{}.
\newblock \showarticletitle{{Construction and Evaluation of a User Experience
  Questionnaire}}. In \bibinfo{booktitle}{\emph{{HCI} and Usability for
  Education and Work, 4th Symposium of the Workgroup Human-Computer Interaction
  and Usability Engineering of the Austrian Computer Society, {USAB} 2008,
  Graz, Austria, November 20-21, 2008. Proceedings}}
  \emph{(\bibinfo{series}{Lecture Notes in Computer Science},
  Vol.~\bibinfo{volume}{5298})}. \bibinfo{publisher}{Springer},
  \bibinfo{pages}{63--76}.
\newblock
\urldef\tempurl%
\url{https://doi.org/10.1007/978-3-540-89350-9\_6}
\showDOI{\tempurl}


\bibitem[Lewis(1982)]%
        {thinking}
\bibfield{author}{\bibinfo{person}{Charles~R. Lewis}.}
  \bibinfo{year}{1982}\natexlab{}.
\newblock \showarticletitle{{Using the "thinking aloud" method in cognitive
  interface design}}. \bibinfo{publisher}{IBM TJ Watson Research Center
  Yorktown Heights, NY}.
\newblock


\bibitem[Ley(2009)]%
        {DBLP:journals/pvldb/Ley09}
\bibfield{author}{\bibinfo{person}{Michael Ley}.}
  \bibinfo{year}{2009}\natexlab{}.
\newblock \showarticletitle{{{DBLP} - Some Lessons Learned}}.
\newblock \bibinfo{journal}{\emph{Proc. {VLDB} Endow.}} \bibinfo{volume}{2},
  \bibinfo{number}{2} (\bibinfo{year}{2009}), \bibinfo{pages}{1493--1500}.
\newblock
\urldef\tempurl%
\url{https://doi.org/10.14778/1687553.1687577}
\showDOI{\tempurl}


\bibitem[Liang et~al\mbox{.}(2021)]%
        {DBLP:conf/semweb/LiangPYZLM21}
\bibfield{author}{\bibinfo{person}{Zhicheng Liang}, \bibinfo{person}{Zixuan
  Peng}, \bibinfo{person}{Xuefeng Yang}, \bibinfo{person}{Fubang Zhao},
  \bibinfo{person}{Yunfeng Liu}, {and} \bibinfo{person}{Deborah~L.
  McGuinness}.} \bibinfo{year}{2021}\natexlab{}.
\newblock \showarticletitle{{BERT-based Semantic Query Graph Extraction for
  Knowledge Graph Question Answering}}. In
  \bibinfo{booktitle}{\emph{Proceedings of the {ISWC} 2021 Posters, Demos and
  Industry Tracks: From Novel Ideas to Industrial Practice co-located with 20th
  International Semantic Web Conference {(ISWC} 2021), Virtual Conference,
  October 24-28, 2021}} \emph{(\bibinfo{series}{{CEUR} Workshop Proceedings},
  Vol.~\bibinfo{volume}{2980})}. \bibinfo{publisher}{CEUR-WS.org}.
\newblock
\urldef\tempurl%
\url{http://ceur-ws.org/Vol-2980/paper379.pdf}
\showURL{%
\tempurl}


\bibitem[Lu(2011)]%
        {10.1093/database/baq036}
\bibfield{author}{\bibinfo{person}{Zhiyong Lu}.}
  \bibinfo{year}{2011}\natexlab{}.
\newblock \showarticletitle{{PubMed and beyond: a survey of web tools for
  searching biomedical literature}}.
\newblock \bibinfo{journal}{\emph{Database}}  \bibinfo{volume}{2011}
  (\bibinfo{date}{01} \bibinfo{year}{2011}).
\newblock
\showISSN{1758-0463}
\urldef\tempurl%
\url{https://doi.org/10.1093/database/baq036}
\showDOI{\tempurl}
\newblock
\shownote{baq036}.


\bibitem[Manola et~al\mbox{.}(2004)]%
        {manola2004rdf}
\bibfield{author}{\bibinfo{person}{Frank Manola}, \bibinfo{person}{Eric
  Miller}, \bibinfo{person}{Brian McBride}, {et~al\mbox{.}}}
  \bibinfo{year}{2004}\natexlab{}.
\newblock \showarticletitle{{RDF primer}}.
\newblock \bibinfo{journal}{\emph{W3C recommendation}} \bibinfo{volume}{10},
  \bibinfo{number}{1-107} (\bibinfo{year}{2004}), \bibinfo{pages}{6}.
\newblock


\bibitem[Priem et~al\mbox{.}(2022)]%
        {2022openalex}
\bibfield{author}{\bibinfo{person}{Jason Priem}, \bibinfo{person}{Heather
  Piwowar}, {and} \bibinfo{person}{Richard Orr}.}
  \bibinfo{year}{2022}\natexlab{}.
\newblock \bibinfo{title}{{OpenAlex: A fully-open index of scholarly works,
  authors, venues, institutions, and concepts}}.
\newblock
\newblock
\urldef\tempurl%
\url{https://doi.org/10.48550/ARXIV.2205.01833}
\showDOI{\tempurl}


\bibitem[Rekabsaz et~al\mbox{.}(2021)]%
        {rekabsaz2021fairnessir}
\bibfield{author}{\bibinfo{person}{Navid Rekabsaz}, \bibinfo{person}{Oleg
  Lesota}, \bibinfo{person}{Markus Schedl}, \bibinfo{person}{Jon Brassey},
  {and} \bibinfo{person}{Carsten Eickhoff}.} \bibinfo{year}{2021}\natexlab{}.
\newblock \showarticletitle{{TripClick: The Log Files of a Large Health Web
  Search Engine}}. In \bibinfo{booktitle}{\emph{{SIGIR} '21: The 44th
  International {ACM} {SIGIR} Conference on Research and Development in
  Information Retrieval, Virtual Event, Canada, July 11-15, 2021}}.
  \bibinfo{publisher}{{ACM}}, \bibinfo{pages}{2507--2513}.
\newblock
\urldef\tempurl%
\url{https://doi.org/10.1145/3404835.3463242}
\showDOI{\tempurl}


\bibitem[Revanth et~al\mbox{.}(2022)]%
        {10.1007/978-981-16-1342-5_21}
\bibfield{author}{\bibinfo{person}{T.~J. Revanth}, \bibinfo{person}{K.~Venkat
  Sai}, \bibinfo{person}{R. Ramya}, \bibinfo{person}{Renusree Chava},
  \bibinfo{person}{V. Sushma}, {and} \bibinfo{person}{B.~S. Ramya}.}
  \bibinfo{year}{2022}\natexlab{}.
\newblock \showarticletitle{{NL2SQL: Natural Language to SQL Query
  Translator}}. In \bibinfo{booktitle}{\emph{Emerging Research in Computing,
  Information, Communication and Applications}}. \bibinfo{publisher}{Springer
  Singapore}, \bibinfo{pages}{267--278}.
\newblock
\showISBNx{978-981-16-1342-5}


\bibitem[Roberts et~al\mbox{.}(2020)]%
        {TREC_PM20}
\bibfield{author}{\bibinfo{person}{Kirk Roberts}, \bibinfo{person}{Dina
  Demner{-}Fushman}, \bibinfo{person}{Ellen~M. Voorhees},
  \bibinfo{person}{Steven Bedrick}, {and} \bibinfo{person}{William~R. Hersh}.}
  \bibinfo{year}{2020}\natexlab{}.
\newblock \showarticletitle{{Overview of the {TREC} 2020 Precision Medicine
  Track}}. In \bibinfo{booktitle}{\emph{Proceedings of the Twenty-Ninth Text
  REtrieval Conference, {TREC} 2020, Virtual Event [Gaithersburg, Maryland,
  USA], November 16-20, 2020}} \emph{(\bibinfo{series}{{NIST} Special
  Publication}, Vol.~\bibinfo{volume}{1266})}. \bibinfo{publisher}{National
  Institute of Standards and Technology {(NIST)}}.
\newblock
\urldef\tempurl%
\url{https://trec.nist.gov/pubs/trec29/papers/OVERVIEW.PM.pdf}
\showURL{%
\tempurl}


\bibitem[Sakai(2007)]%
        {10.1145/1277741.1277756}
\bibfield{author}{\bibinfo{person}{Tetsuya Sakai}.}
  \bibinfo{year}{2007}\natexlab{}.
\newblock \showarticletitle{{Alternatives to Bpref}}. In
  \bibinfo{booktitle}{\emph{Proceedings of the 30th Annual International ACM
  SIGIR Conference on Research and Development in Information Retrieval}}
  (Amsterdam, The Netherlands) \emph{(\bibinfo{series}{SIGIR '07})}.
  \bibinfo{publisher}{Association for Computing Machinery},
  \bibinfo{pages}{71–78}.
\newblock
\showISBNx{9781595935977}
\urldef\tempurl%
\url{https://doi.org/10.1145/1277741.1277756}
\showDOI{\tempurl}


\bibitem[Saleh and Pecina(2019)]%
        {DBLP:conf/ecir/SalehP19}
\bibfield{author}{\bibinfo{person}{Shadi Saleh} {and} \bibinfo{person}{Pavel
  Pecina}.} \bibinfo{year}{2019}\natexlab{}.
\newblock \showarticletitle{{Term Selection for Query Expansion in Medical
  Cross-Lingual Information Retrieval}}. In \bibinfo{booktitle}{\emph{Advances
  in Information Retrieval - 41st European Conference on {IR} Research, {ECIR}
  2019, Cologne, Germany, April 14-18, 2019, Proceedings, Part {I}}}
  \emph{(\bibinfo{series}{Lecture Notes in Computer Science},
  Vol.~\bibinfo{volume}{11437})}. \bibinfo{publisher}{Springer},
  \bibinfo{pages}{507--522}.
\newblock
\urldef\tempurl%
\url{https://doi.org/10.1007/978-3-030-15712-8\_33}
\showDOI{\tempurl}


\bibitem[Tamine and Goeuriot(2022)]%
        {DBLP:journals/csur/TamineG22}
\bibfield{author}{\bibinfo{person}{Lynda Tamine} {and}
  \bibinfo{person}{Lorraine Goeuriot}.} \bibinfo{year}{2022}\natexlab{}.
\newblock \showarticletitle{{Semantic Information Retrieval on Medical Texts:
  Research Challenges, Survey, and Open Issues}}.
\newblock \bibinfo{journal}{\emph{{ACM} Comput. Surv.}} \bibinfo{volume}{54},
  \bibinfo{number}{7} (\bibinfo{year}{2022}), \bibinfo{pages}{146:1--146:38}.
\newblock
\urldef\tempurl%
\url{https://doi.org/10.1145/3462476}
\showDOI{\tempurl}


\bibitem[Voorhees et~al\mbox{.}(2020)]%
        {Voorhees2020TrecCovid}
\bibfield{author}{\bibinfo{person}{Ellen~M. Voorhees}, \bibinfo{person}{Tasmeer
  Alam}, \bibinfo{person}{Steven Bedrick}, \bibinfo{person}{Dina
  Demner{-}Fushman}, \bibinfo{person}{William~R. Hersh}, \bibinfo{person}{Kyle
  Lo}, \bibinfo{person}{Kirk Roberts}, \bibinfo{person}{Ian Soboroff}, {and}
  \bibinfo{person}{Lucy~Lu Wang}.} \bibinfo{year}{2020}\natexlab{}.
\newblock \showarticletitle{{{TREC-COVID:} constructing a pandemic information
  retrieval test collection}}.
\newblock \bibinfo{journal}{\emph{{SIGIR} Forum}} \bibinfo{volume}{54},
  \bibinfo{number}{1} (\bibinfo{year}{2020}), \bibinfo{pages}{1:1--1:12}.
\newblock
\urldef\tempurl%
\url{https://doi.org/10.1145/3451964.3451965}
\showDOI{\tempurl}


\bibitem[Vrandecic(2012)]%
        {DBLP:conf/www/Vrandecic12}
\bibfield{author}{\bibinfo{person}{Denny Vrandecic}.}
  \bibinfo{year}{2012}\natexlab{}.
\newblock \showarticletitle{{Wikidata: a new platform for collaborative data
  collection}}. In \bibinfo{booktitle}{\emph{Proceedings of the 21st World Wide
  Web Conference, {WWW} 2012, Lyon, France, April 16-20, 2012 (Companion
  Volume)}}. \bibinfo{publisher}{{ACM}}, \bibinfo{pages}{1063--1064}.
\newblock
\urldef\tempurl%
\url{https://doi.org/10.1145/2187980.2188242}
\showDOI{\tempurl}


\bibitem[Wang et~al\mbox{.}(2020)]%
        {Wang2020Cord19}
\bibfield{author}{\bibinfo{person}{Lucy~Lu Wang}, \bibinfo{person}{Kyle Lo},
  \bibinfo{person}{Yoganand Chandrasekhar}, \bibinfo{person}{Russell Reas},
  \bibinfo{person}{Jiangjiang Yang}, \bibinfo{person}{Darrin Eide}, {and}
  \bibinfo{person}{al.}} \bibinfo{year}{2020}\natexlab{}.
\newblock \showarticletitle{{CORD-19: The Covid-19 Open Research Dataset}}.
\newblock \bibinfo{journal}{\emph{CoRR}}  \bibinfo{volume}{abs/2004.10706}
  (\bibinfo{year}{2020}).
\newblock
\showeprint[arXiv]{2004.10706}
\urldef\tempurl%
\url{https://arxiv.org/abs/2004.10706}
\showURL{%
\tempurl}


\bibitem[Zenz et~al\mbox{.}(2009)]%
        {10.1016/j.websem.2009.07.005}
\bibfield{author}{\bibinfo{person}{Gideon Zenz}, \bibinfo{person}{Xuan Zhou},
  \bibinfo{person}{Enrico Minack}, \bibinfo{person}{Wolf Siberski}, {and}
  \bibinfo{person}{Wolfgang Nejdl}.} \bibinfo{year}{2009}\natexlab{}.
\newblock \showarticletitle{{From Keywords to Semantic Queries-Incremental
  Query Construction on the Semantic Web}}.
\newblock \bibinfo{journal}{\emph{Web Semant.}} \bibinfo{volume}{7},
  \bibinfo{number}{3} (\bibinfo{date}{sep} \bibinfo{year}{2009}),
  \bibinfo{pages}{166–176}.
\newblock
\showISSN{1570-8268}
\urldef\tempurl%
\url{https://doi.org/10.1016/j.websem.2009.07.005}
\showDOI{\tempurl}


\end{thebibliography}
